\documentclass[journal=jacsat,manuscript=article]{achemso}

\usepackage{chemformula} % Formula subscripts using \ch{}
\usepackage[T1]{fontenc} % Use modern font encodings

\author{Andi Rabia}

\author{Francesco Tumino}

\author{Alberto Milani}

\author{Valeria Russo}

\author{Andrea Li Bassi}

\author{Nicolò Bassi}
\affiliation[Politecnico di Milano]
{Department of Energy, Politecnico di Milano via Ponzio 34/3, I-20133 Milano, Italy.}

\altaffiliation{Current address: Empa, Swiss Federal Laboratories for Materials Science and Technology, 8600 D\"ubendorf, Switzerland.}

\author{Andrea Lucotti}
\affiliation[Politecnico di Milano]
{Dipartimento di Chimica, Materiali e Ingegneria Chimica "G. Natta", Politecnico di Milano, Piazza 
	Leonardo da Vinci 32, 20133 Milano, Italy.}
\author{Simona Achilli}
\affiliation[ Universit\`a degli Studi di Milano]
{ETSF and Dipartimento di Fisica ``Aldo Pontremoli'', Universit\`a degli Studi di Milano, Via Celoria, 16, I-20133 Milano, Italy.}

\author{Guido Fratesi}
\affiliation[ Universit\`a degli Studi di Milano]
{ETSF and Dipartimento di Fisica ``Aldo Pontremoli'', Universit\`a degli Studi di Milano, Via Celoria, 16, I-20133 Milano, Italy.}
\author{Nicola Manini}
\affiliation[ Universit\`a degli Studi di Milano]
{ETSF and Dipartimento di Fisica ``Aldo Pontremoli'', Universit\`a degli Studi di Milano, Via Celoria, 16, I-20133 Milano, Italy.}
\author{Giovanni Onida}
\affiliation[ Universit\`a degli Studi di Milano]
{ETSF and Dipartimento di Fisica ``Aldo Pontremoli'', Universit\`a degli Studi di Milano, Via Celoria, 16, I-20133 Milano, Italy.}
\author{Qiang Sun}
\affiliation[ Tongji University]
{Interdisciplinary Materials Research Center, College of Materials Science and Engineering, Tongji University, Shanghai 201804, P. R. China.}
\altaffiliation{Current address: Materials Genome Institute, Shanghai University, 200444 Shanghai, China.}
\author{Wei Xu}
\affiliation[ Tongji University]
{Interdisciplinary Materials Research Center, College of Materials Science and Engineering, Tongji University, Shanghai 201804, P. R. China.}

\author{Carlo S. Casari}
\email{* carlo.casari@polimi.it}
\phone{ +39 0223996331}
\fax{}
\affiliation[Politecnico di Milano]
{Department of Energy, Politecnico di Milano via Ponzio 34/3, I-20133 Milano, Italy.}

\title[An \textsf{achemso} demo]
{Structural, Electronic, and Vibrational Properties of 2D Graphdiyne-Like Carbon Nanonetwork Synthesized on Au(111): Implications for the Engineering of sp-sp$^2$ Carbon Nanostructures}

\abbreviations{IR,NMR,UV}

\begin{document}

	%\begin{tocentry}
		
	%	\begin{figure}
	%	\includegraphics{TOC.eps}
	%	\end{figure}
%	\end{tocentry}

	\begin{abstract}
		Graphdiyne, atomically-thin 2D carbon nanostructure based on sp-sp$^2$ hybridization, is an appealing system potentially showing outstanding mechanical and optoelectronic properties. Surface-catalyzed coupling of halogenated sp-carbon-based molecular precursors represents a promising bottom-up strategy to fabricate extended 2D carbon systems with engineered structure on metallic substrates. Here, we investigate the atomic-scale structure and electronic and vibrational properties of an extended graphdiyne-like sp-sp$^2$ carbon nanonetwork grown on Au(111) by means of on-surface synthesis. The formation of such 2D nanonetwork at its different stages as a function of the annealing temperature after the deposition is monitored by scanning tunneling microscopy (STM), Raman spectroscopy and combined with density functional theory (DFT) calculations. High-resolution STM imaging and the high sensitivity of Raman spectroscopy to the bond nature provide a unique strategy to unravel the atomic-scale properties of sp-sp$^2$ carbon nanostructures. We show that hybridization between the 2D carbon nanonetwork and the underlying substrate states strongly affects its electronic and vibrational properties, modifying substantially the density of states and the Raman spectrum compared to the free standing system. This opens the way to the modulation of the electronic properties with significant prospects in future applications as active nanomaterials for catalysis, photoconversion and carbon-based nanoelectronics. 
		
	\end{abstract}
	
	\section{Introduction}
	Two-dimensional (2D) carbon systems with mixed sp-sp$^2$ hybridization, i.e., graphyne and graphdiyne, aroused great interest in the scientific community over the last thirty years, as novel 2D carbon structures, \cite{Hirsch_NatMat_2010,baughman1987,Casari_Nanoscale_2016} paving the way for the ultimate goal of fabricating sp-hybridized carbon fragments, whose structural, optical and transport properties were deeply explored \cite{Casari_Nanoscale_2016, Bonardi, Ravagnan, Zanolli}. 
	These systems can form a variety of 2D crystals with different structure, sp/sp$^2$ ratio, density, porosity and have been predicted to possess peculiar electronic properties, such as multiple Dirac
	cones in  graphyne.\cite{Malko_PRL_2012}

	Synthetic chemistry attempts to develop sp-sp$^2$ carbon networks have been carried out by developing efficient methodologies based on a monomer-to-polymer approach. In this respect, one of the most used protocols has been the oxidative acetylenic coupling of macrocyclic carbon-rich precursors\cite{siemsen2000acetylenic}. However, direct coupling of the monomers reported the cross-linking of the chains\cite{bunz1999polyethynylated} which precludes the development of well-defined extended structures. To tackle this issue, Haley\cite{Haley_PureAppChem_2008} proposed the isolation of the reactive acetylene moieties and the assembly of the structure via an intramolecular cyclization approach, obtained through sequential Sonogashira cross-coupling reactions\cite{sonogashira2002development}.
	A graphdiyne film has been for the first time produced by G. Li et al.\cite{Li_2010_GDY} on copper substrate, through cross-coupling reaction, then several works reported the fabrication in form of flakes or powder and tested for potential applications in the field of photocatalysis and nanoelectronics \cite{Yang_2013, Li_Graphyne_molecular_electronics2015, james_2018, Huang_2018, Hui_2019}.
	
	The investigation of the structure of these systems is of fundamental importance, since for a long time graphdiyne has been considered elusive and poor stability of the sp carbon phase is still an important issue opposing the realization of extended and ordered 2D structures.           
	In this context, Raman spectroscopy stands as the election technique to investigate the structure of carbon-based materials, proving to get access to the presence of sp carbon and to the structural properties and local bond order. The characteristic Raman signal of sp carbon and its structure-related behaviour allows a detailed investigation of sp and sp-sp$^2$ carbon systems.\cite{Milani_BeilsteinJ_2015,serafini2020raman}    
	
	Despite the significant advancement in the synthesis and the investigation of these materials, the fabrication of extended graphdiyne monolayers and the imaging of their atomic-scale structure is still a challenge. Scanning tunneling microscopy and spectroscopy (STM/STS) in combination with theoretical modelling has the potential to unravel the atomic-scale structure and to provide a deep insight into the surface electronic properties. The atomic-scale imaging by means of scanning probe techniques usually requires material growth in UHV conditions on atomically flat surfaces and in-situ characterization.   
	In this framework, the rapidly growing on-surface synthesis technique has demonstrated tremendous potential in the high-precision bottom-up construction of low-dimensional carbon nanostructures and in the atomic-scale imaging and characterization by surface science techniques. \cite{zhang2012homo,zhang2015surface,doi:10.1021/ja510292b,doi:10.1002/smll.201603675,gao2013glaser,Brambilla17}. Its great advantage relies on fostering selective chemical reactions between molecular precursors on metal surfaces under ultra-high vacuum (UHV) conditions\cite{xing2019selective,kang2019surface}. To this aim, molecular precursors are designed to favor the adsorption and the subsequent on-surface homocoupling reaction, where the substrate act as a catalytical template triggering the reaction in mild conditions. The targeted nanostructures are directly observed by surface-sensitive techniques, such as STM and atomic force microscopy (AFM)\cite{binning1986scanning}. An unambiguous demonstration of the power of this approach is the synthesis of atomically-precise graphene nanoribbons\cite{cai2010atomically,Ruffieux2016489} (GNRs) with engineered properties\cite{tao2011spatially,dilullo2012molecular,koch2012voltage,llinas2017short,dos2017electric, sun2020massive}. This strategy has been recently developed to synthesize a broad range of novel carbon nanostructures based on sp-hybridization, such as carbon atom wires (CAWs) and 2D extended networks of mixed sp-sp$^2$-hybridization. While the fabrication of sp$^2$ carbon systems is achieved through on-surface Ullman coupling reaction of aryl halides, an efficient strategy for sp carbon nanostructures is represented by on-surface dehalogenative/dehydrogenative homocoupling reaction of precursors functionalized with alkynyl halides\cite{shen2017frontiers,liu2017surface,klappenberger2015surface,Sun2018,Sun_AngewChem_2017,Shu_NatComm_2018,yang2018nanoscale}. 
	However, accessing the nature of molecular bonds and hybridization state represents a key factor to comprehensively understand the properties of the novel sp-sp$^2$ carbon structures. In a previous work we have adopted Raman spectroscopy for the identification of sp-hybridization and investigation of on-surface formation mechanism of sp-sp$^2$ carbon atomic wires on Au(111) substrate\cite{rabia2019scanning}. 
	
	Herein, we report on the nanoscale structure, electronic and vibrational properties of a  carbon monolayer nanonetwork whose structure resembles $\gamma$- or $\alpha$-graphdiyne, while differing by the presence of aromatic rings with 3-fold hydrogen terminated bonds. A 2D carbon nanonetwork based on sp-sp$^2$ hybridization was grown on Au(111) under UHV conditions, exposing to organic molecular precursor with three alkynyl bromide groups\cite{sun2016}. By a combination of  high-resolution STM imaging, Raman spectroscopy and DFT simulations we unveil the structure at different stages of the formation, i.e., from metal organic nanostructure comprising Au adatoms to the pure sp-sp$^2$ carbon nanonetwork obtained after release of Au atoms by thermal annealing. Insight on the 2D surface electronic structure is provided by first principle calculations. Raman spectroscopy and  DFT calculations reveal the C-C vibrational modes of this one-atom thick 2D sp-sp$^2$ carbon nanonetwork on Au(111). The present work provides an example of great potentiality of Raman spectroscopy in studying the atomic-scale structure of novel carbon nanostructures based on sp-sp$^2$-hybridization, e.g. graphyne and graphdiyne, complementing the high-resolution STM imaging. The interaction between the 2D carbon nanonetwork and the underlying Au(111) surface involves a charge transfer and has a fundamental role in modifying both the electronic and vibrational properties of this system with respect to what expected for the free-standing counterpart. The availability of a 2D carbon-based semiconductor on a metal may open new opportunities in the field of catalysis with novel non precious nanomaterials, photoconversion and photovoltaics as well as carbon-based nanoelectronic devices and sensors.

	\section{Results and discussion}
	
\begin{figure*}[ht]
		\centering
		\includegraphics[height=9cm]{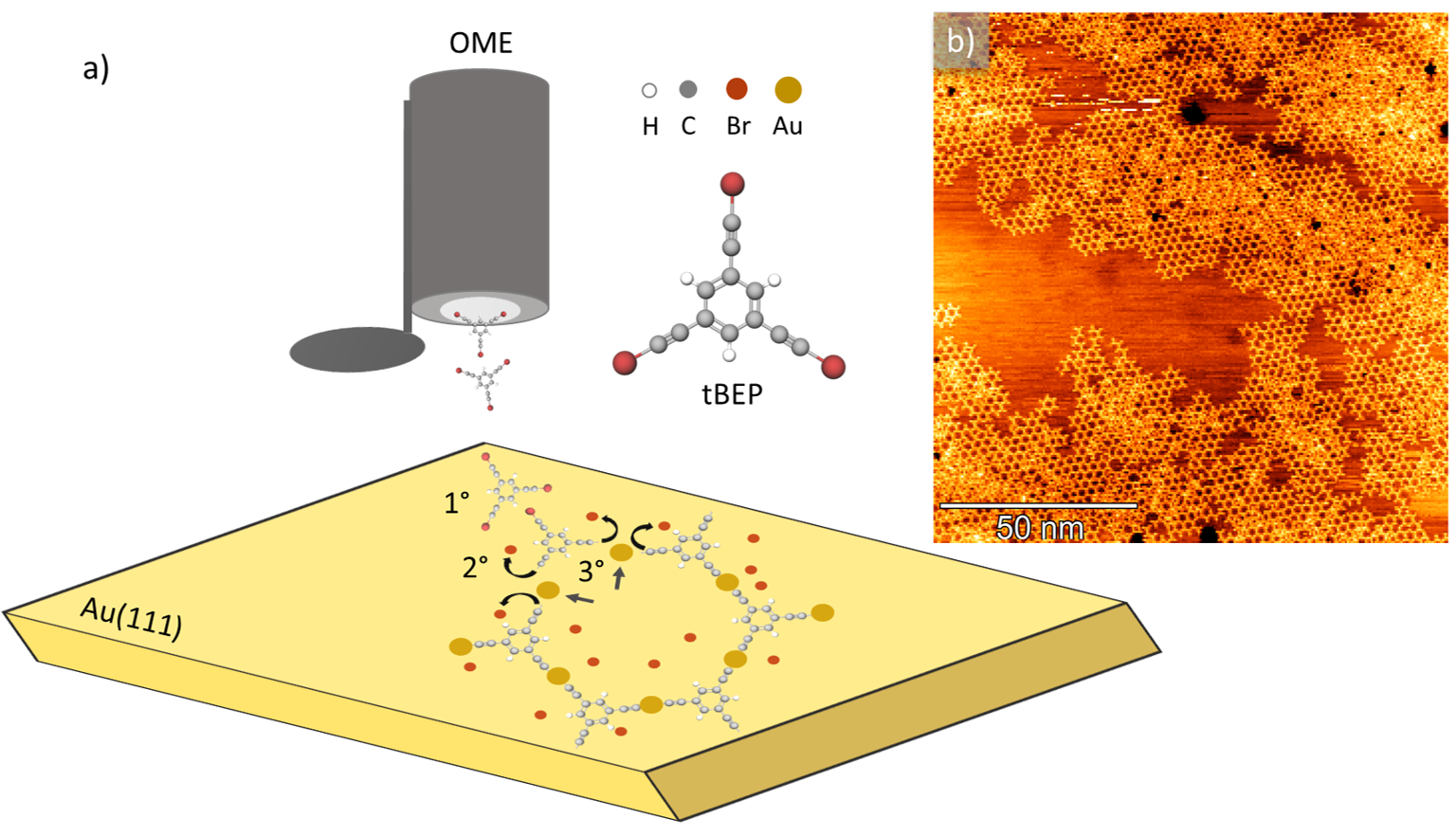}
		\caption{ (a) Scheme showing the deposition of 1,3,5-tris(bromoethynyl)benzene (tBEP) by organic molecular evaporator (OME) and reporting the different steps of on-surface synthesis process on Au(111); adsorption (1°), dehalogenation process (2°) and coupling through Au adatom (3°), (b)  Large-scale STM image showing the sp-sp$^2$ carbon nanonetwork formed after the deposition of tBEP molecular precursor on Au(111) (1.3 V, 0.5 nA).  
		}
		\label{scheme_1}
	\end{figure*}

	The graphdiyne-like 2D sp-sp$^2$ carbon monolayer is obtained on Au(111) after UHV deposition at room temperature of the precursor molecules shown in Fig.~\ref{scheme_1}a. The on-surface synthesis mechanism can be easily sketched in three steps \cite{Sun2018} (see Fig.~\ref{scheme_1}a) which take place simultaneously: i) adsorption of the molecules on Au(111) surface, ii) cleavage of C-Br bonds followed by the diffusion of radicals on the 2D template, iii) coupling of the sp carbon chains through gold adatoms. Following this scheme, after deposition the molecules self-assemble and form an organometallic 2D nanonetwork (i.e., metalated system) based on sp-sp$^2$ carbon, catalyzed by the Au(111) surface. The substrate is subsequently annealed to remove the Au atoms and promote the C-C homocoupling reaction\cite{sun2016} to form an all-carbon nanonetwork. The obtained nanostructure has three sp carbon diacetylenic chains connecting sp$^2$ six-membered aromatic rings in a three-fold organization instead of the six-fold structure of $\gamma$-graphdiyne (the remaining three-fold dangling bonds are saturated by H atoms). The resulting structure shows a large-scale hexagonal organization very similar to the $\alpha$-graphdiyne, except for the presence of sp$^2$ aromatic rings in place of single C atoms.

	In the following sections we present the results of STM and Raman investigation of this system. DFT calculations of the electronic and vibrational properties allow to discuss STM images and Raman spectra and to unveil the fundamental role of the Au(111) surface which affects in a relevant way the density of states and the Raman spectrum.

	%%%%%%%%%%%%%%%-----STM SECTION-----%%%%%%%%%%%%%%%%%%%%%%%%%%%%%%
	\subsection{Structural and electronic properties}\label{sec:stm}
	In the STM image in Fig.~\ref{scheme_1}b we report the formation of sp-sp$^2$ carbon nanonetwork which consists in a honeycomb structure extending on the gold surface. Investigating the coverage obtained on the sample, by acquiring large-scale STM images at different regions of the substrate, we observe an ordered  sp-sp$^2$ carbon nanonetwork over most of the surface. However, there exist fractions of network where the regularity is disrupted by unreacted or possibly degraded molecules deposited as a second layer (Fig.~\ref{scheme_1}b). Boundary regions between sp-sp$^2$ nanonetwork and uncovered Au(111) surface show a disordered morphology (Fig.~\ref{scheme_1}b), probably because of the random endcapping of highly reactive sp carbon chains with substrate atoms.

\begin{figure*}[ht]
		\centering
		\includegraphics[height=4.8 cm]{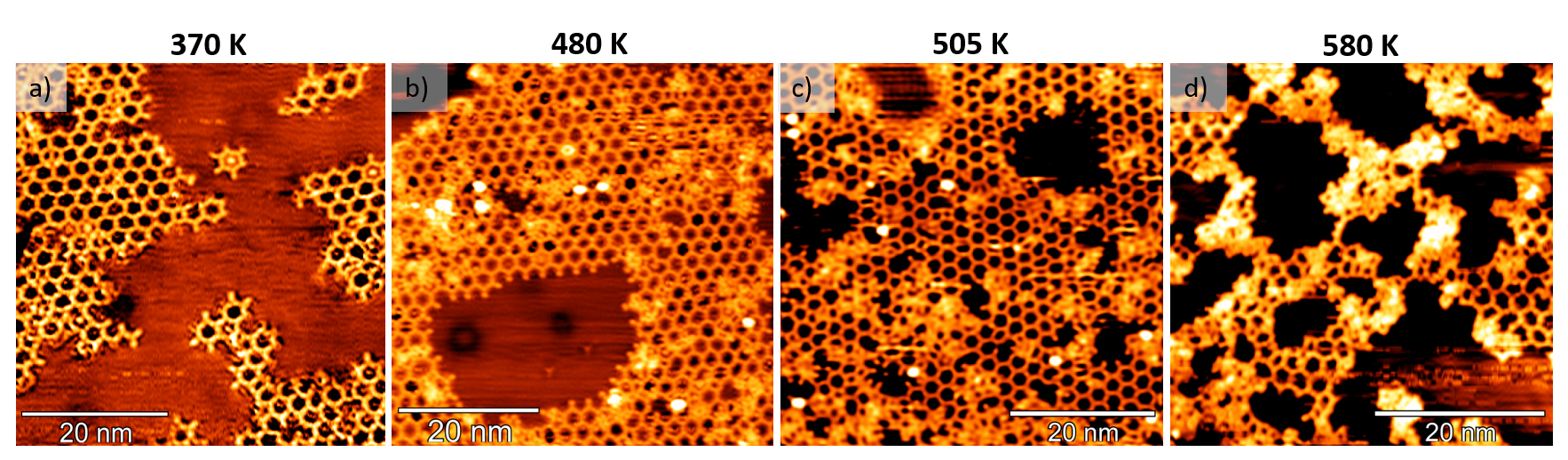}
		\caption{
			RT-STM images showing the morphology of the 2D nanonetwork based on sp-sp$^2$ carbon at different annealing temperatures: (a) a lower coverage observed after annealing the sample at 370 K (1 V, 0.3 nA), (b) the onset of the local disordered phase after annealing at 480 K  (-0.5 V, 0.4 nA), (c) a progressive increase of the disordered phase after annealing at 505 K (-0.8 V, 0.3nA), (d) a pronounced morphological disorder developing after annealing at 580 K (-1.2 V, 0.3 nA).
		}
		\label{morph}
	\end{figure*}

	We studied the effect of thermal annealing on the morphology and structure by acquiring $in-situ$ STM images at RT. To this end, we have annealed the sample to temperatures up to 580~K (Fig.~\ref{morph}) in UHV. STM image of the sample annealed at 370~K (Fig.~\ref{morph}a) report a lower coverage than as-deposited sample (Fig.~\ref{scheme_1}b) and this is an indication of the detachment of the less-strong interacting molecules with the substrate. Then, after increasing the annealing temperature in the range 480 - 580~K (Fig.~\ref{morph}b-d), we observe the formation of an amorphous phase coexisting with the regular 2D nanonetwork. Such patchy structure might be established by the diffusion and thermal activation of the reactive sp carbon atomic chains originating the onset of the disordered phase, 
	as also revealed by the Raman spectra discussed in the next section. The presence of the amorphous phase further increases when the sample is annealed at 580~K, as shown in Fig.~\ref{morph}d, making it difficult to image the surface.
	
\begin{figure}[th]
		\centering
		\includegraphics[height=11cm]{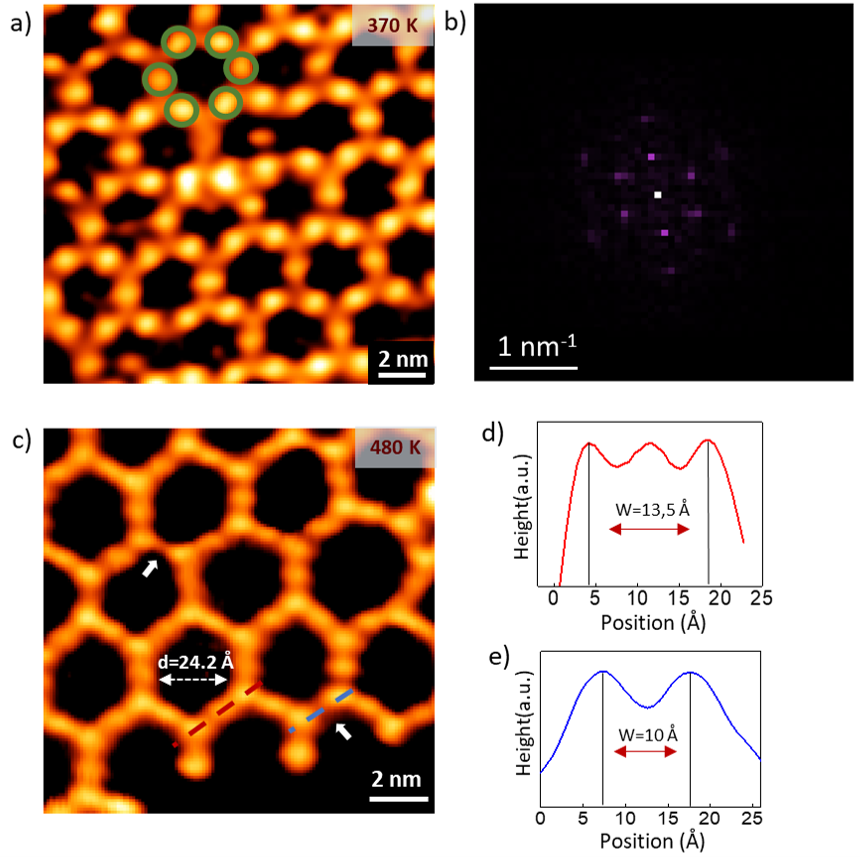}
		\caption{
			High-resolution STM images resolve the structure of sp-sp$^2$ carbon nanonetwork: (a) image of the metalated nanonetwork taken after annealing the sample at 370 K where Au adatoms are circled with green color (0.5 V, 0.3 nA), (b) FFT of (a) showing the hexagonal symmetry, (c) image taken after annealing the sample at 480 K (0.5 V, 0.3 nA) where white arrows points at the missing Au sites, (d) line-profile (red color) on metalated sp carbon chains in (b) showing a peak-to-peak of 13.5 \AA,  (e) line-profile (blue color) on C-C coupled  chains in (b) showing a peak-to-peak length of 10 \AA.
		}
		\label{struct}
	\end{figure}

	In addition to the morphology, we studied the  molecular structure of sp-sp$^2$ carbon nanonetwork by high-resolution STM images and DFT calculations. The sp-sp$^2$ carbon nanonetwork has been obtained through on-surface coupling of tBEP molecules where the substrate catalyzes the reaction by detaching the bromine atoms and substituting them with surface gold adatoms. The crucial role of the surface atoms is pointed out in the STM image acquired on the sample annealed at 370~K (Fig.~\ref{struct}a) where, notably, the color contrast enhances the round and bright protrusions associated to the gold adatoms. In Fig.~\ref{struct}a gold adatoms are circled with green color and they form a Kagomé-like structure. 

	A further support to the structural analysis of metalated  sp-sp$^2$ carbon nanonetwork has been provided by the Fast Fourier Transform (FFT) (Fig.~\ref{struct}b) which reports a hexagonal symmetry and a periodicity of $\sim$ 21 \AA. As the annealing temperature rises to 480 K the hexagons in Fig.~\ref{struct}c show protrusions associated to phenyl rings and intermediate gold atoms, while it is difficult to resolve the atomic structure of short sp carbon atomic chains due to the limited resolution of STM signal. Nevertheless, at 480 K we observe the disappearance of Au atoms in some sp carbon chains which are highlighted by white arrows in Fig.~\ref{struct}c. Indeed, by extracting the topographic profile (Fig.~\ref{struct}d-e) of the metalated and C-C coupled atomic chains (red and blue dashed line in Fig.~\ref{struct}c) we measure the length of the chains around 13.5 and 10 \AA~ respectively. 
	
	%THEORY: Structure
	
\begin{figure}[ht]
		\centering
		\includegraphics[height=10 cm]{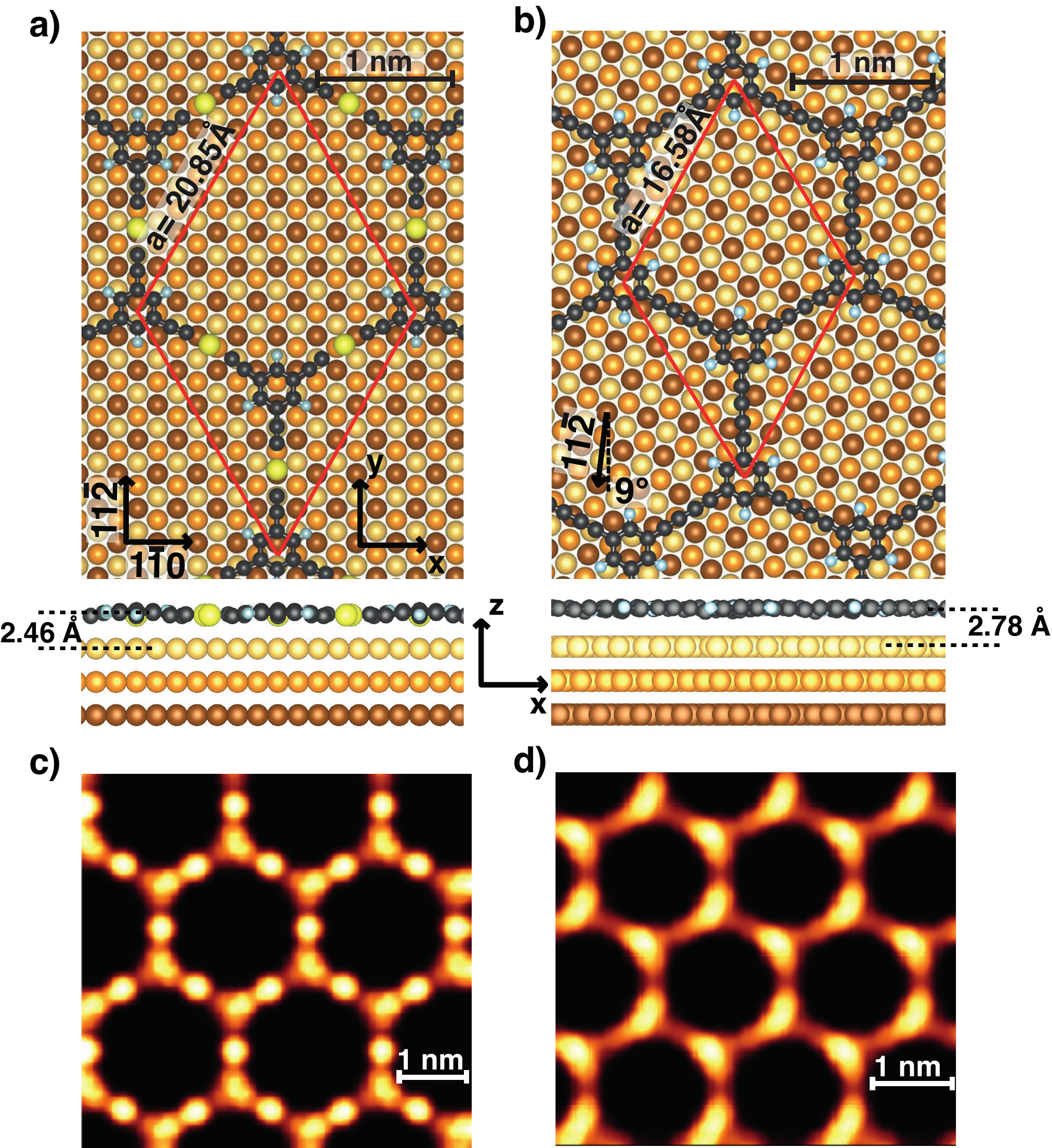}
		\caption{
			Ball-and-stick model of the the metalated (a) and C-C coupled 2D nanonetwork (b) sp-sp$^2$ carbon nanonetwork, where the red arrows represent the unit vectors of the hexagonal lattice. Au atoms are depicted as yellow-orange-brown spheres from the first to the third surface layer. Panel (c) and (d) reports the simulated STM images for the two systems.  
		}
		\label{struct-theo}
	\end{figure}

	% --------------------
	Figure \ref{struct-theo} reports the ball-and-stick model of the metalated and C-C coupled 2D nanonetworks on Au(111) adopted in the calculations. 
	The size of the unit cell was fixed to match as close as possible the periodicity extracted from the experimental STM-line profile and from the FFT, imposing a commensurate matching of the overlayer with the substrate (as also required in calculations with periodic boundary conditions). In particular for the metalated case the experimental periodicity is $\sim$ 21 \AA~and fits with a $7 \times 7$ unit cell of Au(111), corresponding to 20.85 \AA~(where the theoretical lattice constant $a_{th}=2.97$ \AA~has been used). With such values, the sp-sp$^2$ 2D nanonetwork on Au(111) is compressed by 0.5$\%$ with compared to the theoretical periodicity calculated for the freestanding nanonetwork. In this model the sp carbon chains are aligned along the $\left[11\bar2\right]$ direction. For the C-C coupled system the periodicity extracted from STM is about 17 \AA. A commensurate structure with similar lattice constant for the molecular overlayer on Au(111) can be obtained by rotating the overlayer by $8.9 ^\circ$ relative to the $\left[11\bar1\right]$ direction of the substrate in this configuration. The periodicity of the hexagonal lattice amount to 16.58 \AA~($\sim5.6~a_{th}$) accounting for an expansion of 0.7 $\%$ with respect to the freestanding 2D carbon nanonetwork.

	Upon geometrical relaxation the sp-sp$^2$ metalated 2D nanonetwork exhibits a mild bending of two sides of the hexagon (on the left in Fig.~\ref{struct-theo}) probably due to the small compression of the overlayer. 
	The bond lengths of the adsorbed polymer are 125 pm, and 199 pm for the C-C triple bond and C-Au, respectively. Moreover, the interaction between the Au adatom and the substrate induces a corrugation of the molecular nanonetwork leading to a smaller distance between the Au atom in the chain and the substrate (2.46 \AA) compared to the average phenyl-substrate distance (2.70 \AA). These data are in agreement with previous theoretical findings for the analogous 1D system \cite{rabia2019scanning}.
	
	The C-C coupled system is decoupled from the substrate and displays a negligible out-of-plane and in-plane distortion and a larger distance from the substrate that amount to 2.78 \AA. The C-C single and triple bond lengths are 136 pm and 125 pm, respectively.

	The theoretical simulation of the STM images, reported in Fig.~\ref{struct-theo}, are in agreement with the experiments, showing bright protrusion in correspondence of Au atoms in the chain for the metalated nanonetwork.
	The theory predicts also a slightly enhanced STM contrast on the phenyl groups with respect to the attached ``arms''.

	For the all-carbon case the model shows a bright contrast of the atomic carbon chains along the rotated $\left [11\bar2\right]$ direction and a less intense signal in proximity to one end of the four oblique chains.
	This effect can be related to the different alignment of the atomic carbon chains with the underlying substrate in the adopted structural model. The resulting modulation of the heights of carbon atoms gives rise to a geometrical effect in the simulation.

	% ELECTRONIC PROPERTIES  -- THEORY --

\begin{figure*}[ht]
		\centering
		\includegraphics[height=7cm]{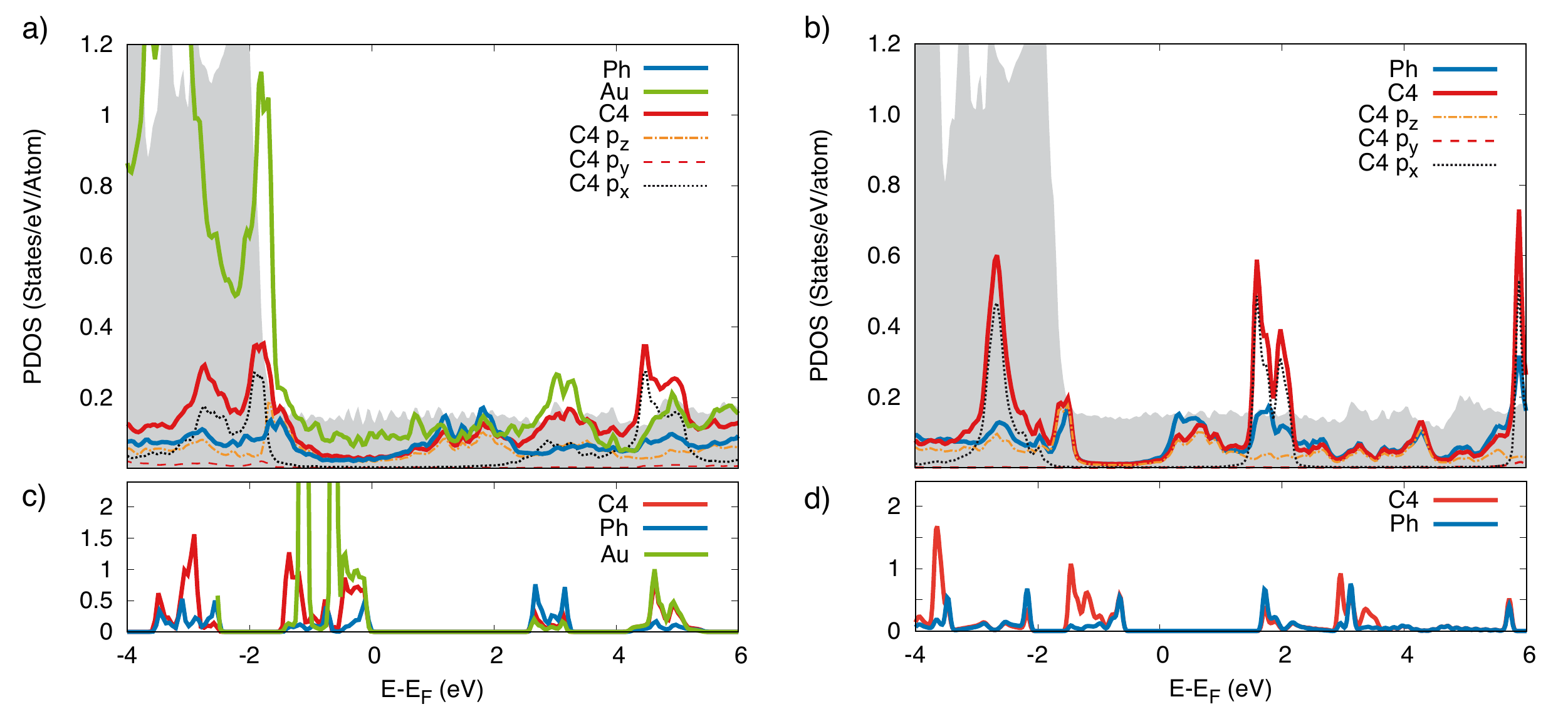}
		\caption{Top panels: Density of States projected on different groups and atoms for the metalated (a) and C-C coupled 2D nanonetwork (b) 2D nanonetwork. The shaded area represents the PDOS of the bare Au(111) substrate. (c) and (d): PDOS of the freestanding systems metalated and C-C coupled systems, respectively.   
		}
		\label{PDOS}
	\end{figure*}

	In order to investigate the variation of the electronic properties induced by the adsorption of precursors, formation of the 2D nanonetwork on the surface and by the removal of the gold atom upon annealing, we have analyzed the projected density of states (PDOS) integrated in the whole Brillouin zone for both the 2D nanonetworks supported on the Au(111) surface, in comparison with the freestanding system. Figure~\ref{PDOS} reports the integrated PDOS of the phenyl groups of the molecule and of the sp chain aligned with the $\left [11\bar2\right]$ direction. The latter is also resolved with respect to the magnetic quantum number ($m$) of the p orbitals.
	
	In the metalated 2D nanonetwork the states of the gold atom in the sp chain (green line in Fig.~\ref{PDOS}a) display a relevant superposition with the $d$-band of the Au(111) surface (gray shaded area).
	The chain-substrate interaction is also the driving force in the charge transfer process that determines an increase of electronic charge ($+0.68~e$) on the molecule with respect to the freestanding system.
	The energy separation related to the HOMO-LUMO gap is still visible, even though the hybridization with the states of the substrate induces a metal behavior in the 2D overlayer.

	The analysis of the $m$-components of the PDOS allows us to identify the contribution of p states of the sp chain along the different directions.
	In making the projection, we consider the average of the C atoms of a specific sp chain and align the $y$ axis along the chain direction. Indeed, considering all the sp chains would result in averaging in-plane.
	Hence, p$_\mathrm{x}$, p$_\mathrm{y}$, and p$_\mathrm{z}$ orbitals are aligned along $\left[1\bar10\right]$, $\left [11\bar2\right]$, and [111] directions, respectively.

	In particular the states closer to the Fermi level have p$_\mathrm{z}$ symmetry and form out-of-plane $\pi$-states. Other $\pi$-states (orthogonal to the sp chain axis) now lying parallel to the surface, are formed by the p$_\mathrm{x}$; those are more separated in energy, being located below $-2$~eV and above $2$~eV. Noticeably, p$_\mathrm{x}$ states are sharper in the PDOS than p$_\mathrm{z}$ states, due to a smaller hybridization with the Au(111) orbitals. Similar observations could be drawn for a linear sp-sp$^2$ system\cite{frat2018}.
	Conversely, p$_\mathrm{y}$ states are sit farther from the Fermi level both in the occupied (valence band) and unoccupied (conduction band) part, as expected from their $\sigma$ character and direction along the chain axis.
	
	Upon removal of the Au adatom, the 2D nanonetwork is decoupled from the substrate and maintains the energy gap around the Fermi level, although reduced with respect to the freestanding nanonetwork ($\approx 1.6$~eV versus $\approx 2.1$~eV).
	The lack of the metallic character related to the hybridization with the substrate is associated to a lower electronic charge on the carbon nanonetwork with respect to the metalated case and comparable to that of the freestanding layer ($-0.012~e$ on the supported 2D carbon nanonetwork).
	The ordering of the different p components is the same observed for the metallic system. In particular the occupied p states of the sp carbon chain lie in correspondence to the d band of Au(111).
	
	The theoretical calculation can also provide a simulation of STS images, as the DOS at the $\bar{\Gamma}$ point, which is representative of the states with the maximum interaction with the tip due to their slow decay in vacuum \cite{Donati2009} (see Supporting Information).

	DFT calculations reveal that the interaction between the 2D carbon nanonetwork and the underlying Au(111) surface consistently affects the electronic properties in comparison with the free standing carbon nanostructure. As detailed above and in the Supporting Information, the metalated nanostructure acquires a metallic character due to a consistent charge transfer while the graphdiyne-like nanonetwork displays a modified gap with respect to the free-standing system. In these sp carbon based systems peculiar structure-dependent conjugation effects and relevant electron-phonon coupling lead to a strong relationship between electronic and vibrational properties. The effect of the interacting Au(111) surface on the vibrational properties of 2D carbon system is addressed by comparing experimental Raman spectrum with DFT calculations, as discussed in the next section.     
	
	%%%%%%%%%%%%%%%-----RAMAN SECTION-----%%%%%%%%%%%%%%%%%%%%%%%%%%%%%%
	\subsection{Raman spectroscopy and vibrational properties}\label{sec:Raman}
	The Raman spectra of the precursor and of the graphdyine-like 2D carbon nanonetwork are here discussed and interpreted in view of DFT calculations of the Raman response. To provide a reliable interpretation of the experimental trends we performed DFT calculations on the precursor molecule and on a molecular model describing the final 2D all-carbon nanostructure grown on the substrate.

	\begin{figure}[th]
		\centering
		\includegraphics[height=10cm]{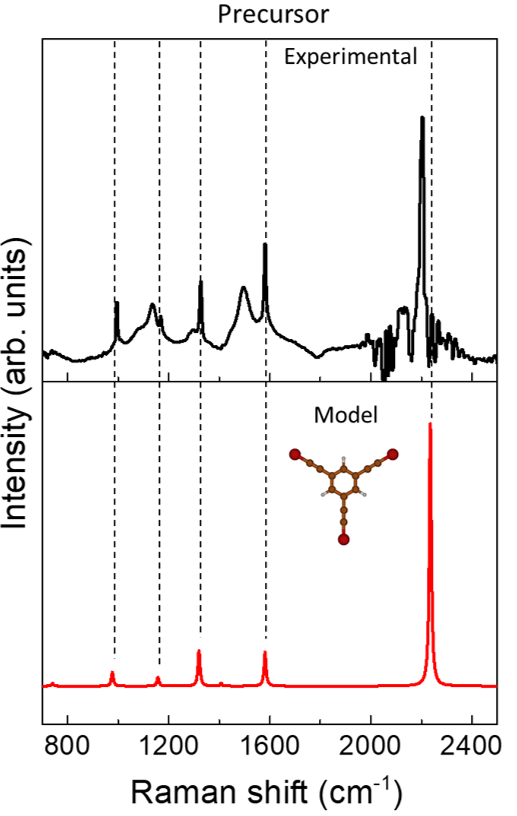}
		\caption{DFT calculation (red) and experimental FT-Raman spectrum (black) of the molecular precursor.
			The DFT spectra have been rescaled to match the CC stretching mode of the phenyl ring at 1581 cm$^{-1}$.
		}
		\label{Raman_1}
	\end{figure}
	
	%%%%%% RAMAN OF PRECURSOR  %%%%%%%%%%%%%%%%%
	We start by discussing the spectrum of the molecular precursor (see Fig.~\ref{Raman_1}) measured in powder form, prior to sublimation. In particular, FT-Raman spectroscopy has been employed to acquire the Raman spectrum of the organic molecule and to avoid the strong luminescence background.
	By comparing the FT-Raman spectrum (black) of the initial molecule in Fig.~\ref{Raman_1} with the simulated spectrum of the precursor (red), a very good agreement is found, demonstrating the accuracy of the adopted level of theory in providing an accurate interpretation of the Raman response. The experimental spectrum shows three main spectral features: i) the  sp carbon stretching mode at 2203 cm$^{-1}$, ii) the peak associated to the sp$^{2}$ carbon in the phenyl rings at 1581 cm$^{-1}$ and iii) few weak peaks in low-wavenumber region (950-1350 cm$^{-1}$) which show a very good correspondence with the peaks predicted in the theoretical spectrum. On the other hand, the two broad bands observed at about 1500 cm$^{-1}$ and 1150 cm$^{-1}$ can be attributed to impurities, also possibly due to degradation processes. DFT calculations allow us to assign the peaks in detail, starting from the peak at 2236 cm$^{-1}$ (computed at an unscaled wavenumber of 2319~cm$^{-1}$) which is due to CC stretching mode of the triple bond. The peak at 1581 cm$^{-1}$ (computed at an unscaled 1641 cm$^{-1}$) is attributed to CC stretching mode localized on the phenyl groups, while the bands at lower wavenumber values (1370, 1201 and 1014 cm$^{-1}$ unscaled values) are associated respectively to CC stretching of the phenyl coupled to CC stretching on the three branches, and different combination of CC and CH bending modes on the phenyl also coupled to the contribution of CBr stretching. 
	
\begin{figure}[th]
		\centering
		\includegraphics[height=12cm]{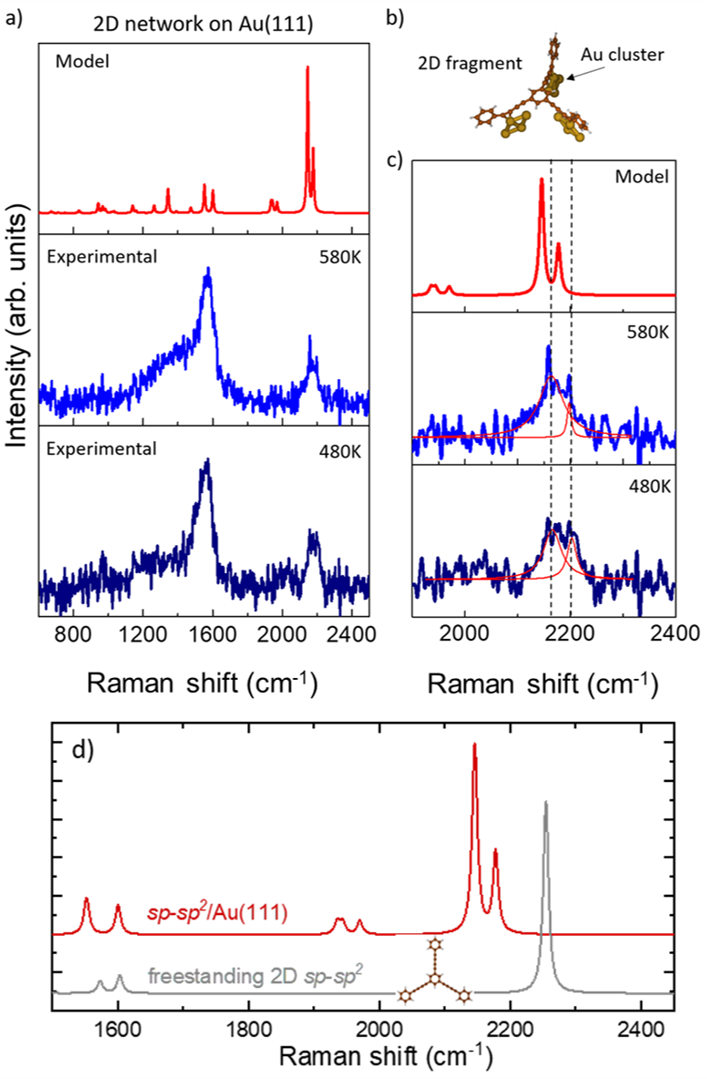}
		\caption{(a) Ex-situ Raman spectra of the 2D nanonetwork on Au(111) annealed at different temperatures and DFT calculations carried out on the molecular model shown in (b). (b) Complex comprised of fragment of the final 2D nanonetwork and Au cluster to mimick the interaction of the 2D nanonetwork with the substrate. (c) Magnification of the spectral range relevant to sp carbon (1900~cm$^{-1}$- 2400~cm$^{-1}$). d) The calculated spectrum of the graphdyine nanonetwork on Au(111) of panels (a) and (b) is compared with the freestanding case.    
		}
		\label{Raman_2}
	\end{figure}

	%%%%%% RAMAN OF sample  %%%%%%%%%%%%%%%%%
	Ex-situ Raman spectra of the 2D nanonetwork on Au(111) annealed at 480~K and 580~K have been acquired to follow the evolution as a function of the temperature (see Fig.~\ref{Raman_2}). The experimental spectra of the 2D nanonetwork on Au(111) (Fig.~\ref{Raman_2}) appear to be significantly modified compared to the spectrum of the precursor molecule (Fig.~\ref{Raman_1}). Indeed, the weak signals at lower frequency (950-1350 cm$^{-1}$) disappear while the peaks associated to sp carbon and sp$^{2}$ carbon in the phenyl ring become broader. By fitting the sp carbon band with two gaussians (Fig.~\ref{Raman_2}c) (one centered at 2200 cm$^{-1}$ and the other one at 2165 cm$^{-1}$) we observe that the contribution of the lower frequency peak (2165 cm$^{-1}$) increases after annealing the sample at 580~K. Concerning the spectral region associated to the sp$^{2}$ carbon in the phenyl ring, we observe a broad band shifted to lower-frequency (from 1581 cm$^{-1}$ to 1565 cm$^{-1}$) compared to the peak of the precursor molecule. The broadening of this band is further enhanced at 580~K probably due to the increased disorder observed in STM images and discussed below. 
	
	We computed the Raman spectrum of the all-carbon 2D nanonetwork on Au(111) by adopting the simplified model in which small Au$_4$ clusters are interacting with the sp domains of a fragment of the 2D nanonetwork, as shown in Fig.~\ref{Raman_2}b (see Supplementary Information Fig.~S3 for an extended discussion). DFT calculations predict a double peak associated to the sp carbon chains at high frequency range, namely at 2145 cm$^{-1}$ and 2178 cm$^{-1}$. This doublet falls at lower wavenumber with respect to the sp carbon peak predicted for the precursor at 2236 cm$^{-1}$ and is associated now to collective stretching modes involving the different sp carbon domains, as described for sp carbon chain by the effective conjugation coordinate (ECC) model\cite{Milani_BeilsteinJ_2015,Casari_Nanoscale_2016}. The modulation of their frequency is due to the interactions with the gold clusters, which can affect the stretching force constants of the CC bond most involved in the interaction with gold, as also found in similar systems\cite{rabia2019scanning}. As a result, the whole collective CC stretching  mode of the sp carbon chains is perturbed by the non-bonding interactions, and the peaks are shifted to lower wavenumber with respect to those predicted for the unperturbed precursor molecule. This result is in agreement with what we observe in the experimental spectra, where the sp carbon band can be fitted by two peaks at slightly lower wavenumber with respect to the precursor molecule. Therefore, DFT calculations allow us to give an interpretation to the Raman spectra indicating the formation of the 2D nanostructure on Au(111) as also observed by STM imaging. Morevover, DFT calculations predict other two weak peaks at 1939 and 1969 cm$^{-1}$. The normal mode associated to these peaks can be described as CC stretching vibrations mainly localized on those CC bonds which are more strongly interacting with the gold cluster. Due to their weak intensity these peaks are difficult to observe in the experimental spectra, they could be possibly corresponding to the weak bump at 2025 cm$^{-1}$ barely observable in both spectra.
	
	The comparison with DFT calculations allows us to identify the Raman spectrum of the 2D nanostructure on Au(111). The two main peaks at slightly lower wavenumber (2165 and 2200 cm$^{-1}$ ) than the precursor molecule provides a clear signature of the occurrence of homocoupling of the molecules on the substrate. Although STM images provide a clear evidence of the metalated 2D nanostructure formed right after the deposition of the molecules on Au(111), we were not able to record an ex-situ Raman spectrum due to low signal-to-noise ratio. Such behavior was observed also in the case of 1D nanostructures based on sp-sp$^{2}$ carbon studied in previous work\cite{rabia2019scanning} and we think that annealing the system in UHV can improve the signal-to-noise ratio, possibly due to higher stability of the whole nanonetwork after exposure to the atmosphere. However, from STM images we notice a wide temperature window for the homocoupling reaction and the metalated domains are still present after annealing at 480~K and much less at 580~K. The progressive transformation of the structure with the thermal annealing is reflected in the Raman spectra where the peaks at 2165 and 2200 cm$^{-1}$ exhibit a change in intensity and width with the temperature (Fig.~\ref{Raman_2}.(c)). Lastly, the peak at 1565 cm$^{-1}$, which shows broadening at 580~K with the onset of a pronounced shoulder, can be interpreted as a superposition of the signal coming from the aromatic rings and a combination of D and G bands typical of sp$^{2}$-carbon amorphous phase, in agreement with the disordered phase observed in the STM image in Fig.~\ref{morph}(d).
	
	The interaction with Au turns out to modify significantly the vibrational properties and the Raman spectrum. By comparing DFT calculations of the sp-sp$^{2}$ system on Au(111) with the freestanding case, three main effects can be seen: i) the ECC peak of sp-carbon stretching mode redshifts; ii) the ECC peak splits in a doublet and iii) a new band at about 2000 cm$^{-1}$ is present which is split in two peaks as well. No substantial differences are seen in the peaks at about 1600 cm$^{-1}$ related to the sp$^{2}$-carbon stretching in the phenyl rings. 
	
	The previous work carried out by us \cite{rabia2019scanning} reported the combined STM and Raman analysis of 1D sp-sp$^2$ polymeric carbon wires on Au(111), observing a similar softening of the ECC mode and the appearance of an additional peak at about 2000 cm$^{-1}$ due to the interaction of the carbon system with Au. These two systems have in common Raman peaks in the same three regions, i.e., the sp carbon fingerprint ($2150-2200$ cm$^{-1}$), the sp$^{2}$ carbon stretching in the phenyl ring (1550-1600 cm$^{-1}$) and a peculiar band at about 2000 cm$^{-1}$. However, the graphdiyne-like nanostructure shows systematic splitting of the peaks in the above mentioned regions compared to the 1D polymeric system. The similarity of the spectra could be expected since the precursor molecules are very similar (the present system has just one more bromoethynyl group) and the CC stretching modes are mainly localized in the $sp$-carbon chains and in the phenyl ring, respectively. Instead, the splitting of the peaks could be attributed to the interaction with gold which decouples vibrational modes due to the breaking of the local symmetry.
	
	Apart from the splitting found in the Raman spectra as a consequence of the interactions with the Au surface, it is also interesting to compare the DFT computed Raman spectra shown in Fig. \ref{Raman_2}.d for the free-standing model with previous calculations of free-standing 2D crystal of $\gamma$-graphdyine, as reported in the paper by Zhang et al.\cite{Zhang_JPCC2016-graphyne} and Serafini et al.\cite{serafini2020raman}. In all the cases, as expected, intense bands are predicted above 2000 cm$^{-1}$ and at 1581 cm$^{-1}$, consistently with the existence of sp carbon domains connected by phenyl units. On the other hands, two distinct and intense bands are predicted for $\gamma$-graphdyine (2142 and 2221 cm$^{-1}$ in Zhang et al. \cite{Zhang_JPCC2016-graphyne}, 2276 cm$^{-1}$ and 2335 cm$^{-1}$  in Serafini et al. \cite{serafini2020raman}) while only one is found here (2339 cm$^{-1}$ , unscaled value). The present system presents however a different structure, possessing a three-fold symmetry, instead of the six-fold symmetry of $\gamma$-graphdiyne. This structural difference affects the vibrational spectra of the two systems, generating different topology-dependent spectral patterns. In the context of graphdiyne-like materials, these results  show the relevance of Raman spectroscopy as a characterization technique suitable to investigate and discriminate between sp-sp$^{2}$ hybrid carbon systems characterized by different topologies.

	\subsection{Conclusions}
	
	We characterized the atomic-scale structure and the vibrational properties of carbon sp-sp$^2$ 2D nanonetwork on Au(111) by means of STM and Raman spectroscopy. In particular, high-resolution STM imaging combined with Raman spectroscopy allowed us to identify the metalated nanostructure formed right after the deposition, and the subsequent formation of sp-sp$^2$ 2D nanonetwork upon annealing in UHV. Raman spectra reported the redshift and the splitting of the sp carbon fingerprint (ECC modes) and the activation of new Raman frequencies not originally present in freestanding nanostructure. The overall redshift of the sp carbon modes and the new Raman peak observed at significantly lower frequencies is attributed to bond softening promoted by the interaction with gold atoms.

	The DFT calculations of the electronic properties of both metalated and C-C coupled nanonetwork on Au(111) show a strong contribution of the substrate states and shifting of molecular states. Indeed, the metalated nanonetwork exhibits a broad peak at Fermi level related to the charge transfer from the substrate to the 2D layer, while for the C-C coupled system the interaction with the substrate is less important but still modifies the gap and the vibrational properties. Although we obtained sp-sp$^2$ 2D nanonetworks with different domain size, we did not observe any variation of the properties with the size, since the sp conjugation is limited in the short sp carbon chain and the phenyl ring. A larger extension of the $\pi$-electron conjugation could be achieved through the control of the oligomeric unit by a proper design of the molecular precursor, finely tuning the properties of the sp-sp$^2$ 2D nanonetwork.

	Although different research groups have proposed novel graphdiyne-like nanostructures obtained by on-surface synthesis \cite{Sun2018, Fan20152484, klappenberger2015surface, klappenberger2018functionalized}, the characterization of their vibrational properties by Raman spectroscopy is lacking. On the line opened by the present work, UHV-STM and Raman spectroscopy represent a powerful and non-destructive combination of techniques for the characterization of novel sp carbon systems, such as graphdiyne produced by means of on-surface synthesis on metal surfaces. These systems can display unique properties resulting from the interaction with the metal surface whose presence is a key factor to catalyse the formation of the 2D layer, itself. The control of such interaction through an accurate selection of the systems could open to the design of novel 2D carbon materials beyond graphene.  
	The choice of the precursor molecule plays an important role in determining the final properties of the nanostructure \cite{Gao_2019}. For instance, the content and architecture of the linear sp carbon structures defines the sp/sp$^2$ ratio and pore-size. The chemical functionalization of the initial molecule, such as with halogens, tailors the interaction with substrate, which catalyzes the on-surface synthesis and modifies the electronic and vibrational properties of the resulting system.
	
	Compared with $\gamma$-graphdiyne structure, our carbon nanonetwork shows a three-fold connection of sp$^{2}$ six-membered aromatic rings through diacetylenic linkages and the remaining three-fold dangling bonds are saturated by H atoms. Accordingly, it possesses a uniformly larger pore-size (six times larger surface area) and lower sp/sp$^2$ ratio. The latter suggests a higher band-gap energy compared to $\gamma$-graphdiyne, as confirmed by the prediction of about 1.2 eV for $\gamma$-graphdiyne\cite{serafini2020raman} 
	with respect to 2.1 eV for the structure reported here. In addition, the electronic properties of this 2D graphdiyne-like carbon nanonetwork are strongly affected by the interaction with the metallic substrate resulting in a decrease of the band-gap to about 1.6 eV. The possibility to modulate the pore-size and electronic properties of 2D graphdiyne-like nanomaterial represents a great potential for future applications in catalysis \cite{Huang_2018, Zuo_2018} and nanoelectronics \cite{james_2018}.
	
	\section{Experimental and computational details}
	
	The experiments were carried out in two interconnected  ultra-high vacuum chambers with a base pressure of less than $5\times10^{-11}$ mbar. The first experimental set-up was used for the STM  measurements while the second setup was devoted to sample preparation. 
	
	The Au(111) surface was prepared by repeated cycles of Ar$^+$ ion sputtering, followed by annealing at 720 K. The \textit{herringbone reconstruction} characterizes the clean Au(111) surface with atomically-flat terraces as confirmed by STM measurements. The molecular precursor, i. e.  1,3,5-tris(bromoethynyl)benzene (tBEP), was thoroughly degassed prior to the deposition onto the cleaned Au(111) surface kept at room-temperature (RT). The tBEP precursor was thermally sublimated onto Au(111) surface by means of an organic molecular evaporator (OME) at 304 K. STM measurements were performed by employing an Omicron variable-temperature scanning tunneling microscope. STM images were taken in constant-current mode at room temperature, with a chemically etched tungsten tip. The bias for tunneling current is applied to the sample and typical voltages are from $-1.5$ to $+1.5$ V and tunneling current in the range 0.2-1.0 nA.
	
	Density functional theory calculations have been performed by adopting standard norm-conserving pseudopotentials and an atomic-orbitals basis set which includes double-zeta and polarization orbitals, as implemented in the SIESTA package \cite{Sole02}.
	
	The exchange and correlation potential was treated with the generalized gradient approximation (GGA-PBE)
	\cite{PBE}
	and the molecule-surface Van der Waals interaction was introduced
	via a DFT-D2 Grimme potential.
	\cite{Grimme}.
	The mesh cutoff for the real space grid was set to 450 Ry and a $3\times3$ and $4\times4$ sampling of the Brillouin zone was adopted for the metalated and C-C coupled system, respectively, corresponding to approximately a $21\times21$ grid in the primitive Brillouin zone of Au(111).
	
	A five times denser grid was used for the calculation of the density of states (DOS).
	The molecular layers were fully relaxed until the forces reached
	the tolerance value of 0.04 eV~\AA$^{-1}$. The substrate atoms were
	kept fixed to the coordinates of the unrelaxed ideal clean
	Au(111) surface, neglecting the $22\times\sqrt{3}$ reconstruction. Along
	the $z$-direction we consider six gold layers, with an interposed
	vacuum of 38~\AA.
	The STM simulations were performed in a Tersoff–Hamann
	approach, assuming a constant density of states for the tip.
	We integrated the electronic density of the empty states in an
	energy interval 0.5 eV just above the Fermi level. STM images were simulated at constant-distance of 3~{\AA}- from the surface and applying a 2~{\AA}-wide
	Gaussian spatial broadening to the electronic density to mimic
	finite experimental resolution.

	MicroRaman measurements were conducted, $ex situ$, using a Renishaw InVia spectrometer coupled with an Argon laser (514.5 nm). With the power set at 5 mW, we have acquired 100 spectra for each sample aiming at an adequate signal-to-noise ratio. At this excitation wavelength we have a fluorescent background coming from Au(111) surface and concurrent with Raman signal of the sp-sp$^2$ carbon atomic sheet. The background has been removed by subtracting the signal acquired on the pristine clean Au(111) surface under the same experimental conditions. The Raman spectrum of the molecular precursor in powder form was obtained by FT-Raman ($NICOLET$  $NEXUS$ $NXR$  $9650$ ) with 1064 nm excitation wavelength. 
	
	The simulation of the Raman spectrum of the precursor and of the carbon nanonetwork have been carried out by applying the approach successfully adopted in a previous paper\cite{rabia2019scanning} on a similar system, using finite-dimension molecular models which properly describe the real system. Modelling the present system is indeed far from being trivial. An extended computational investigation has been carried out to this aim as reported in detail in the Supporting Information.
	After the calculation of the molecular precursor, DFT calculations have been performed for an isolated fragment of the 2D crystal interacting with three Au clusters, in order to model the interaction expected between the Au surface and the sp domains \cite{rabia2019scanning}. This model is reported in section 3.2. The Gaussian09 package \cite{g09} has been used to carry out these calculations, adopting the PBE0 functional and a cc-pVTZ basis set for C and H and ECP60MDF pseudopotential together with VTZ basis set for Au \cite{ECP_Au}. The computed vibrational frequencies have been rescaled by a factor of 0.963, determined by adjusting the phenyl stretching mode predicted by DFT at 1641 cm$^{-1}$ to the value 1581 cm$^{-1}$ observed experimentally  for the precursor.

	\begin{acknowledgement}

	\end{acknowledgement}
	Authors acknowledge A. Mele and C. Gambarotti of Politecnico di Milano 
	for their support in checking the stability of precursor molecules by gaschromatography and mass spectrometry. The authors also aknowledge  CINECA under the ISCRA initiative (project HP10C3S9Z0) and Red Espa{\~n}ola de Supercomputacion (FI-2020-1-0014) for the availability of high performance computing resources and support. AR, FT, AM, VR, ALB, NB, AL, CSC acknowledge funding from the European Research Council (ERC) under the European Union’s Horizon 2020 research and innovation program ERC-Consolidator Grant (ERC CoG 2016 EspLORE grant agreement No. 724610, website: www.esplore.polimi.it).

	
	\begin{suppinfo}
		
		\begin{itemize}
			\item Supporting Information: Theoretical simulation of STS spectra and a detailed discussion on methods and models adopted for the computation of Raman spectra of both molecular precursor and 2D system on Au(111).
			
		\end{itemize}
		
	\end{suppinfo}

	\bibliography{achemso-demo}

\providecommand{\latin}[1]{#1}
\makeatletter
\providecommand{\doi}
  {\begingroup\let\do\@makeother\dospecials
  \catcode`\{=1 \catcode`\}=2 \doi@aux}
\providecommand{\doi@aux}[1]{\endgroup\texttt{#1}}
\makeatother
\providecommand*\mcitethebibliography{\thebibliography}
\csname @ifundefined\endcsname{endmcitethebibliography}
  {\let\endmcitethebibliography\endthebibliography}{}
\begin{mcitethebibliography}{58}
\providecommand*\natexlab[1]{#1}
\providecommand*\mciteSetBstSublistMode[1]{}
\providecommand*\mciteSetBstMaxWidthForm[2]{}
\providecommand*\mciteBstWouldAddEndPuncttrue
  {\def\EndOfBibitem{\unskip.}}
\providecommand*\mciteBstWouldAddEndPunctfalse
  {\let\EndOfBibitem\relax}
\providecommand*\mciteSetBstMidEndSepPunct[3]{}
\providecommand*\mciteSetBstSublistLabelBeginEnd[3]{}
\providecommand*\EndOfBibitem{}
\mciteSetBstSublistMode{f}
\mciteSetBstMaxWidthForm{subitem}{(\alph{mcitesubitemcount})}
\mciteSetBstSublistLabelBeginEnd
  {\mcitemaxwidthsubitemform\space}
  {\relax}
  {\relax}

\bibitem[Hirsch({2010})]{Hirsch_NatMat_2010}
Hirsch,~A. {The era of Carbon allotropes}. \emph{{Nat. Mater.}}
  \textbf{{2010}}, \emph{{9}}, {868--871}\relax
\mciteBstWouldAddEndPuncttrue
\mciteSetBstMidEndSepPunct{\mcitedefaultmidpunct}
{\mcitedefaultendpunct}{\mcitedefaultseppunct}\relax
\EndOfBibitem
\bibitem[Baughman \latin{et~al.}(1987)Baughman, Eckhardt, and
  Kertesz]{baughman1987}
Baughman,~R.~H.; Eckhardt,~H.; Kertesz,~M. Structure‐property predictions for
  new planar forms of carbon: Layered phases containing sp$^2$ and sp atoms.
  \emph{J. Chem. Phys.} \textbf{1987}, \emph{87}, 6687--6699\relax
\mciteBstWouldAddEndPuncttrue
\mciteSetBstMidEndSepPunct{\mcitedefaultmidpunct}
{\mcitedefaultendpunct}{\mcitedefaultseppunct}\relax
\EndOfBibitem
\bibitem[Casari \latin{et~al.}({2016})Casari, Tommasini, Tykwinski, and
  Milani]{Casari_Nanoscale_2016}
Casari,~C.~S.; Tommasini,~M.; Tykwinski,~R.~R.; Milani,~A. {Carbon-atom wires:
  1-D systems with tunable properties}. \emph{{Nanoscale}} \textbf{{2016}},
  \emph{{8}}, {4414--4435}\relax
\mciteBstWouldAddEndPuncttrue
\mciteSetBstMidEndSepPunct{\mcitedefaultmidpunct}
{\mcitedefaultendpunct}{\mcitedefaultseppunct}\relax
\EndOfBibitem
\bibitem[Bonardi \latin{et~al.}({2015})Bonardi, Achilli, Tantardini, and
  Martinazzo]{Bonardi}
Bonardi,~P.; Achilli,~S.; Tantardini,~G.~F.; Martinazzo,~R. {Electron transport
  in carbon wires in contact with Ag electrodes: a detailed first principles
  investigation}. \emph{{Phys. Chem. Chem. Phys.}} \textbf{{2015}},
  \emph{{17}}, {18413--18425}\relax
\mciteBstWouldAddEndPuncttrue
\mciteSetBstMidEndSepPunct{\mcitedefaultmidpunct}
{\mcitedefaultendpunct}{\mcitedefaultseppunct}\relax
\EndOfBibitem
\bibitem[Ravagnan \latin{et~al.}({2009})Ravagnan, Manini, Cinquanta, Onida,
  Sangalli, Motta, Devetta, Bordoni, Piseri, and Milani]{Ravagnan}
Ravagnan,~L.; Manini,~N.; Cinquanta,~E.; Onida,~G.; Sangalli,~D.; Motta,~C.;
  Devetta,~M.; Bordoni,~A.; Piseri,~P.; Milani,~P. {Effect of axial torsion on
  sp carbon atomic wires}. \emph{{Phys. Rev. Lett.}} \textbf{{2009}},
  \emph{{102}}, {245502}\relax
\mciteBstWouldAddEndPuncttrue
\mciteSetBstMidEndSepPunct{\mcitedefaultmidpunct}
{\mcitedefaultendpunct}{\mcitedefaultseppunct}\relax
\EndOfBibitem
\bibitem[Zanolli \latin{et~al.}({2010})Zanolli, Onida, and Charlier]{Zanolli}
Zanolli,~Z.; Onida,~G.; Charlier,~J. {Quantum spin transport in carbon chains}.
  \emph{{ACS Nano}} \textbf{{2010}}, \emph{{4}}, {5174}\relax
\mciteBstWouldAddEndPuncttrue
\mciteSetBstMidEndSepPunct{\mcitedefaultmidpunct}
{\mcitedefaultendpunct}{\mcitedefaultseppunct}\relax
\EndOfBibitem
\bibitem[Malko \latin{et~al.}({2012})Malko, Neiss, Vines, and
  Goerling]{Malko_PRL_2012}
Malko,~D.; Neiss,~C.; Vines,~F.; Goerling,~A. {Competition for Graphene:
  Graphynes with Direction-Dependent Dirac Cones}. \emph{{Phys. Rev. Lett.}}
  \textbf{{2012}}, \emph{{108}}, {045443}\relax
\mciteBstWouldAddEndPuncttrue
\mciteSetBstMidEndSepPunct{\mcitedefaultmidpunct}
{\mcitedefaultendpunct}{\mcitedefaultseppunct}\relax
\EndOfBibitem
\bibitem[Siemsen \latin{et~al.}(2000)Siemsen, Livingston, and
  Diederich]{siemsen2000acetylenic}
Siemsen,~P.; Livingston,~R.~C.; Diederich,~F. Acetylenic coupling: a powerful
  tool in molecular construction. \emph{Angew. Chem. Int. Ed.} \textbf{2000},
  \emph{39}, 2632--2657\relax
\mciteBstWouldAddEndPuncttrue
\mciteSetBstMidEndSepPunct{\mcitedefaultmidpunct}
{\mcitedefaultendpunct}{\mcitedefaultseppunct}\relax
\EndOfBibitem
\bibitem[Bunz \latin{et~al.}(1999)Bunz, Rubin, and
  Tobe]{bunz1999polyethynylated}
Bunz,~U.~H.; Rubin,~Y.; Tobe,~Y. Polyethynylated cyclic $\pi$-systems:
  scaffoldings for novel two and three-dimensional carbon networks. \emph{Chem.
  Soc. Rev.} \textbf{1999}, \emph{28}, 107--119\relax
\mciteBstWouldAddEndPuncttrue
\mciteSetBstMidEndSepPunct{\mcitedefaultmidpunct}
{\mcitedefaultendpunct}{\mcitedefaultseppunct}\relax
\EndOfBibitem
\bibitem[Haley({2008})]{Haley_PureAppChem_2008}
Haley,~M.~M. {Synthesis and properties of annulenic subunits of graphyne and
  graphdiyne nanoarchitectures}. \emph{{Pure Appl. Chem.}} \textbf{{2008}},
  \emph{{80}}, {519--532}\relax
\mciteBstWouldAddEndPuncttrue
\mciteSetBstMidEndSepPunct{\mcitedefaultmidpunct}
{\mcitedefaultendpunct}{\mcitedefaultseppunct}\relax
\EndOfBibitem
\bibitem[Sonogashira(2002)]{sonogashira2002development}
Sonogashira,~K. Development of Pd--Cu catalyzed cross-coupling of terminal
  acetylenes with sp2-carbon halides. \emph{J. Organomet. Chem.} \textbf{2002},
  \emph{653}, 46--49\relax
\mciteBstWouldAddEndPuncttrue
\mciteSetBstMidEndSepPunct{\mcitedefaultmidpunct}
{\mcitedefaultendpunct}{\mcitedefaultseppunct}\relax
\EndOfBibitem
\bibitem[Li \latin{et~al.}(2010)Li, Li, Liu, Guo, Li, and Zhu]{Li_2010_GDY}
Li,~G.; Li,~Y.; Liu,~H.; Guo,~Y.; Li,~Y.; Zhu,~D. Architecture of graphdiyne
  nanoscale films. \emph{Chem. Commun.} \textbf{2010}, \emph{46},
  3256--3258\relax
\mciteBstWouldAddEndPuncttrue
\mciteSetBstMidEndSepPunct{\mcitedefaultmidpunct}
{\mcitedefaultendpunct}{\mcitedefaultseppunct}\relax
\EndOfBibitem
\bibitem[Yang \latin{et~al.}(2013)Yang, Liu, Wen, Tang, Zhao, Li, and
  Wang]{Yang_2013}
Yang,~N.; Liu,~Y.; Wen,~H.; Tang,~Z.; Zhao,~H.; Li,~Y.; Wang,~D. Photocatalytic
  Properties of Graphdiyne and Graphene Modified TiO2: From Theory to
  Experiment. \emph{ACS Nano} \textbf{2013}, \emph{7}, 1504--1512\relax
\mciteBstWouldAddEndPuncttrue
\mciteSetBstMidEndSepPunct{\mcitedefaultmidpunct}
{\mcitedefaultendpunct}{\mcitedefaultseppunct}\relax
\EndOfBibitem
\bibitem[Li \latin{et~al.}(2015)Li, Smeu, Rives, Maraval, Chauvin, and
  Ratner]{Li_Graphyne_molecular_electronics2015}
Li,~Z.; Smeu,~M.; Rives,~A.; Maraval,~V.; Chauvin,~R.; Ratner,~E.,~Mark
  A.and~Borguet Towards graphyne molecular electronics. \emph{Nat. Commun.}
  \textbf{2015}, \emph{6}, 6321\relax
\mciteBstWouldAddEndPuncttrue
\mciteSetBstMidEndSepPunct{\mcitedefaultmidpunct}
{\mcitedefaultendpunct}{\mcitedefaultseppunct}\relax
\EndOfBibitem
\bibitem[James \latin{et~al.}(2018)James, John, Owais, Myakala, Chandra~Shekar,
  Choudhuri, and Swathi]{james_2018}
James,~A.; John,~C.; Owais,~C.; Myakala,~S.~N.; Chandra~Shekar,~S.;
  Choudhuri,~J.~R.; Swathi,~R.~S. Graphynes: indispensable nanoporous
  architectures in carbon flatland. \emph{RSC Adv.} \textbf{2018}, \emph{8},
  22998--23018\relax
\mciteBstWouldAddEndPuncttrue
\mciteSetBstMidEndSepPunct{\mcitedefaultmidpunct}
{\mcitedefaultendpunct}{\mcitedefaultseppunct}\relax
\EndOfBibitem
\bibitem[Huang \latin{et~al.}(2018)Huang, Li, Wang, Xue, Zuo, Liu, and
  Li]{Huang_2018}
Huang,~C.; Li,~Y.; Wang,~N.; Xue,~Y.; Zuo,~Z.; Liu,~H.; Li,~Y. Progress in
  Research into 2D Graphdiyne-Based Materials. \emph{Chem. Rev.} \textbf{2018},
  \emph{118}, 7744--7803, PMID: 30048120\relax
\mciteBstWouldAddEndPuncttrue
\mciteSetBstMidEndSepPunct{\mcitedefaultmidpunct}
{\mcitedefaultendpunct}{\mcitedefaultseppunct}\relax
\EndOfBibitem
\bibitem[Hui \latin{et~al.}(2019)Hui, Xue, Yu, Liu, Fang, Xing, Huang, and
  Li]{Hui_2019}
Hui,~L.; Xue,~Y.; Yu,~H.; Liu,~Y.; Fang,~Y.; Xing,~C.; Huang,~B.; Li,~Y. Highly
  Efficient and Selective Generation of Ammonia and Hydrogen on a
  Graphdiyne-Based Catalyst. \emph{J. Am. Chem. Soc.} \textbf{2019},
  \emph{141}, 10677--10683, PMID: 31149825\relax
\mciteBstWouldAddEndPuncttrue
\mciteSetBstMidEndSepPunct{\mcitedefaultmidpunct}
{\mcitedefaultendpunct}{\mcitedefaultseppunct}\relax
\EndOfBibitem
\bibitem[Milani \latin{et~al.}({2015})Milani, Tommasini, Russo, Bassi, Lucotti,
  Cataldo, and Casari]{Milani_BeilsteinJ_2015}
Milani,~A.; Tommasini,~M.; Russo,~V.; Bassi,~A.~L.; Lucotti,~A.; Cataldo,~F.;
  Casari,~C.~S. {Raman spectroscopy as a tool to investigate the structure and
  electronic properties of Carbon-atom wires}. \emph{{Beilstein J.
  Nanotechnol.}} \textbf{{2015}}, \emph{{6}}, {480--491}\relax
\mciteBstWouldAddEndPuncttrue
\mciteSetBstMidEndSepPunct{\mcitedefaultmidpunct}
{\mcitedefaultendpunct}{\mcitedefaultseppunct}\relax
\EndOfBibitem
\bibitem[Serafini \latin{et~al.}(2020)Serafini, Milani, Tommasini, Castiglioni,
  and Casari]{serafini2020raman}
Serafini,~P.; Milani,~A.; Tommasini,~M.; Castiglioni,~C.; Casari,~C.~S. Raman
  and IR spectra of graphdiyne nanoribbons. \emph{Phys. Rev. Mater.}
  \textbf{2020}, \emph{4}, 014001\relax
\mciteBstWouldAddEndPuncttrue
\mciteSetBstMidEndSepPunct{\mcitedefaultmidpunct}
{\mcitedefaultendpunct}{\mcitedefaultseppunct}\relax
\EndOfBibitem
\bibitem[Zhang \latin{et~al.}(2012)Zhang, Kep{\v{c}}ija, Kleinschrodt, Diller,
  Fischer, Papageorgiou, Allegretti, Bj{\"o}rk, Klyatskaya, Klappenberger,
  Ruben, and Barth]{zhang2012homo}
Zhang,~Y.-Q.; Kep{\v{c}}ija,~N.; Kleinschrodt,~M.; Diller,~K.; Fischer,~S.;
  Papageorgiou,~A.~C.; Allegretti,~F.; Bj{\"o}rk,~J.; Klyatskaya,~S.;
  Klappenberger,~F.; Ruben,~M.; Barth,~J.~V. Homo-coupling of terminal alkynes
  on a noble metal surface. \emph{Nat. Commun.} \textbf{2012}, \emph{3},
  1286\relax
\mciteBstWouldAddEndPuncttrue
\mciteSetBstMidEndSepPunct{\mcitedefaultmidpunct}
{\mcitedefaultendpunct}{\mcitedefaultseppunct}\relax
\EndOfBibitem
\bibitem[Zhang \latin{et~al.}(2015)Zhang, Lin, Sun, Chen, Zagranyarski,
  Aghdassi, Duhm, Li, Zhong, Li, Müllen, Fuchs, and Chi]{zhang2015surface}
Zhang,~H.; Lin,~H.; Sun,~K.; Chen,~L.; Zagranyarski,~Y.; Aghdassi,~N.;
  Duhm,~S.; Li,~Q.; Zhong,~D.; Li,~Y.; Müllen,~K.; Fuchs,~H.; Chi,~L.
  On-Surface Synthesis of Rylene-Type Graphene Nanoribbons. \emph{J. Am. Chem.
  Soc.} \textbf{2015}, \emph{137}, 4022--4025, PMID: 25775004\relax
\mciteBstWouldAddEndPuncttrue
\mciteSetBstMidEndSepPunct{\mcitedefaultmidpunct}
{\mcitedefaultendpunct}{\mcitedefaultseppunct}\relax
\EndOfBibitem
\bibitem[Basagni \latin{et~al.}(2015)Basagni, Sedona, Pignedoli, Cattelan,
  Nicolas, Casarin, and Sambi]{doi:10.1021/ja510292b}
Basagni,~A.; Sedona,~F.; Pignedoli,~C.~A.; Cattelan,~M.; Nicolas,~L.;
  Casarin,~M.; Sambi,~M. Molecules–Oligomers–Nanowires–Graphene
  Nanoribbons: A Bottom-Up Stepwise On-Surface Covalent Synthesis Preserving
  Long-Range Order. \emph{J. Am. Chem. Soc.} \textbf{2015}, \emph{137},
  1802--1808\relax
\mciteBstWouldAddEndPuncttrue
\mciteSetBstMidEndSepPunct{\mcitedefaultmidpunct}
{\mcitedefaultendpunct}{\mcitedefaultseppunct}\relax
\EndOfBibitem
\bibitem[Pham \latin{et~al.}(2017)Pham, Tran, Nguyen, and
  St\"ohr]{doi:10.1002/smll.201603675}
Pham,~T.~A.; Tran,~B.~V.; Nguyen,~M.-T.; St\"ohr,~M. Chiral-Selective Formation
  of 1D Polymers Based on Ullmann-Type Coupling: The Role of the Metallic
  Substrate. \emph{Small} \textbf{2017}, \emph{13}, 1603675\relax
\mciteBstWouldAddEndPuncttrue
\mciteSetBstMidEndSepPunct{\mcitedefaultmidpunct}
{\mcitedefaultendpunct}{\mcitedefaultseppunct}\relax
\EndOfBibitem
\bibitem[Gao \latin{et~al.}(2013)Gao, Wagner, Zhong, Franke, Studer, and
  Fuchs]{gao2013glaser}
Gao,~H.-Y.; Wagner,~H.; Zhong,~D.; Franke,~J.-H.; Studer,~A.; Fuchs,~H. Glaser
  coupling at metal surfaces. \emph{Angew. Chem. Int. Ed.} \textbf{2013},
  \emph{52}, 4024--4028\relax
\mciteBstWouldAddEndPuncttrue
\mciteSetBstMidEndSepPunct{\mcitedefaultmidpunct}
{\mcitedefaultendpunct}{\mcitedefaultseppunct}\relax
\EndOfBibitem
\bibitem[{Brambilla} \latin{et~al.}(2017){Brambilla}, {Picone}, {Giannotti},
  {Calloni}, {Berti}, {Bussetti}, {Achilli}, {Fratesi}, {Trioni}, {Vinai},
  {Torelli}, {Panaccione}, {Duò}, {Finazzi}, and {Ciccacci}]{Brambilla17}
{Brambilla},~A.; {Picone},~A.; {Giannotti},~D.; {Calloni},~A.; {Berti},~G.;
  {Bussetti},~G.; {Achilli},~S.; {Fratesi},~G.; {Trioni},~M.~I.; {Vinai},~G.;
  {Torelli},~P.; {Panaccione},~G.; {Duò},~L.; {Finazzi},~M.; {Ciccacci},~F.
  Enhanced Magnetic Hybridization of a Spinterface through Insertion of a
  Two-Dimensional Magnetic Oxide Layer. \emph{Nano Lett.} \textbf{2017},
  \emph{17}, 7440\relax
\mciteBstWouldAddEndPuncttrue
\mciteSetBstMidEndSepPunct{\mcitedefaultmidpunct}
{\mcitedefaultendpunct}{\mcitedefaultseppunct}\relax
\EndOfBibitem
\bibitem[Xing \latin{et~al.}(2019)Xing, Zhang, Fei, Zhao, Zhang, Lin, Zhao, Ju,
  Xu, Fan, Zhu, Ma, and Shi]{xing2019selective}
Xing,~S.; Zhang,~Z.; Fei,~X.; Zhao,~W.; Zhang,~R.; Lin,~T.; Zhao,~D.; Ju,~H.;
  Xu,~H.; Fan,~J.; Zhu,~J.; Ma,~Y.-q.; Shi,~Z. Selective on-surface covalent
  coupling based on metal-organic coordination template. \emph{Nat. Commun.}
  \textbf{2019}, \emph{10}, 1--10\relax
\mciteBstWouldAddEndPuncttrue
\mciteSetBstMidEndSepPunct{\mcitedefaultmidpunct}
{\mcitedefaultendpunct}{\mcitedefaultseppunct}\relax
\EndOfBibitem
\bibitem[Kang and Xu(2019)Kang, and Xu]{kang2019surface}
Kang,~F.; Xu,~W. On-Surface Synthesis of One-Dimensional Carbon-Based
  Nanostructures via C- X and C- H Activation Reactions. \emph{Chem. Phys.
  Chem} \textbf{2019}, \emph{20}, 2251--2261\relax
\mciteBstWouldAddEndPuncttrue
\mciteSetBstMidEndSepPunct{\mcitedefaultmidpunct}
{\mcitedefaultendpunct}{\mcitedefaultseppunct}\relax
\EndOfBibitem
\bibitem[Binning and Rohrer(1986)Binning, and Rohrer]{binning1986scanning}
Binning,~G.; Rohrer,~H. \emph{Scanning Tunneling Microscopy}; Springer, 1986;
  pp 40--54\relax
\mciteBstWouldAddEndPuncttrue
\mciteSetBstMidEndSepPunct{\mcitedefaultmidpunct}
{\mcitedefaultendpunct}{\mcitedefaultseppunct}\relax
\EndOfBibitem
\bibitem[Cai \latin{et~al.}(2010)Cai, Ruffieux, Jaafar, Bieri, Braun,
  Blankenburg, Muoth, Seitsonen, Saleh, Feng, M\"{u}llen, and
  Fasel]{cai2010atomically}
Cai,~J.; Ruffieux,~P.; Jaafar,~R.; Bieri,~M.; Braun,~T.; Blankenburg,~S.;
  Muoth,~M.; Seitsonen,~A.~P.; Saleh,~M.; Feng,~X.; M\"{u}llen,~K.; Fasel,~R.
  Atomically precise bottom-up fabrication of graphene nanoribbons.
  \emph{Nature} \textbf{2010}, \emph{466}, 470--473\relax
\mciteBstWouldAddEndPuncttrue
\mciteSetBstMidEndSepPunct{\mcitedefaultmidpunct}
{\mcitedefaultendpunct}{\mcitedefaultseppunct}\relax
\EndOfBibitem
\bibitem[Ruffieux \latin{et~al.}(2016)Ruffieux, Wang, Yang, Sanchez-Sanchez,
  Liu, Dienel, Talirz, Shinde, Pignedoli, Passerone, Dumslaff, Feng, M\"ollen,
  and Fasel]{Ruffieux2016489}
Ruffieux,~P.; Wang,~S.; Yang,~B.; Sanchez-Sanchez,~C.; Liu,~J.; Dienel,~T.;
  Talirz,~L.; Shinde,~P.; Pignedoli,~C.; Passerone,~D.; Dumslaff,~T.; Feng,~X.;
  M\"ollen,~K.; Fasel,~R. On-surface synthesis of graphene nanoribbons with
  zigzag edge topology. \emph{Nature} \textbf{2016}, \emph{531}, 489--492\relax
\mciteBstWouldAddEndPuncttrue
\mciteSetBstMidEndSepPunct{\mcitedefaultmidpunct}
{\mcitedefaultendpunct}{\mcitedefaultseppunct}\relax
\EndOfBibitem
\bibitem[Tao \latin{et~al.}(2011)Tao, Jiao, Yazyev, Chen, Feng, Zhang, Capaz,
  Tour, Zettl, Louie, Dai, and Crommie]{tao2011spatially}
Tao,~C.; Jiao,~L.; Yazyev,~O.~V.; Chen,~Y.-C.; Feng,~J.; Zhang,~X.;
  Capaz,~R.~B.; Tour,~J.~M.; Zettl,~A.; Louie,~S.~G.; Dai,~H.; Crommie,~M.~F.
  Spatially resolving edge states of chiral graphene nanoribbons. \emph{Nat.
  Phys.} \textbf{2011}, \emph{7}, 616--620\relax
\mciteBstWouldAddEndPuncttrue
\mciteSetBstMidEndSepPunct{\mcitedefaultmidpunct}
{\mcitedefaultendpunct}{\mcitedefaultseppunct}\relax
\EndOfBibitem
\bibitem[DiLullo \latin{et~al.}(2012)DiLullo, Chang, Baadji, Clark,
  Kl\"{o}ckner, Prosenc, Sanvito, Wiesendanger, Hoffmann, and
  Hla]{dilullo2012molecular}
DiLullo,~A.; Chang,~S.-H.; Baadji,~N.; Clark,~K.; Kl\"{o}ckner,~J.-P.;
  Prosenc,~M.-H.; Sanvito,~S.; Wiesendanger,~R.; Hoffmann,~G.; Hla,~S.-W.
  Molecular kondo chain. \emph{Nano letters} \textbf{2012}, \emph{12},
  3174--3179\relax
\mciteBstWouldAddEndPuncttrue
\mciteSetBstMidEndSepPunct{\mcitedefaultmidpunct}
{\mcitedefaultendpunct}{\mcitedefaultseppunct}\relax
\EndOfBibitem
\bibitem[Koch \latin{et~al.}(2012)Koch, Ample, Joachim, and
  Grill]{koch2012voltage}
Koch,~M.; Ample,~F.; Joachim,~C.; Grill,~L. Voltage-dependent conductance of a
  single graphene nanoribbon. \emph{Nat. Nanotechnol.} \textbf{2012}, \emph{7},
  713\relax
\mciteBstWouldAddEndPuncttrue
\mciteSetBstMidEndSepPunct{\mcitedefaultmidpunct}
{\mcitedefaultendpunct}{\mcitedefaultseppunct}\relax
\EndOfBibitem
\bibitem[Llinas \latin{et~al.}(2017)Llinas, Fairbrother, Barin, Shi, Lee, Wu,
  Choi, Braganza, Lear, Kau, Choi, Chen, Pedramrazi, Dumslaff, Narita, Feng,
  M\"{u}llen, Fischer, Zettl, Ruffieux, Yablonovitch, Fasel, and
  Bokor]{llinas2017short}
Llinas,~J.~P. \latin{et~al.}  Short-channel field-effect transistors with
  9-atom and 13-atom wide graphene nanoribbons. \emph{Nat. Commun.}
  \textbf{2017}, \emph{8}, 1--6\relax
\mciteBstWouldAddEndPuncttrue
\mciteSetBstMidEndSepPunct{\mcitedefaultmidpunct}
{\mcitedefaultendpunct}{\mcitedefaultseppunct}\relax
\EndOfBibitem
\bibitem[dos Santos \latin{et~al.}(2017)dos Santos, Mota, Rivelino, and
  Gueorguiev]{dos2017electric}
dos Santos,~R.~B.; Mota,~F. d.~B.; Rivelino,~R.; Gueorguiev,~G.~K.
  Electric-field control of spin-polarization and semiconductor-to-metal
  transition in carbon-atom-chain devices. \emph{J. Chem. Phys. C.}
  \textbf{2017}, \emph{121}, 26125--26132\relax
\mciteBstWouldAddEndPuncttrue
\mciteSetBstMidEndSepPunct{\mcitedefaultmidpunct}
{\mcitedefaultendpunct}{\mcitedefaultseppunct}\relax
\EndOfBibitem
\bibitem[Sun \latin{et~al.}(2020)Sun, Gröning, Overbeck, Braun, Perrin,
  Borin~Barin, El~Abbassi, Eimre, Ditler, Daniels, Meunier, Pignedoli, Calame,
  Fasel, and Ruffieux]{sun2020massive}
Sun,~Q.; Gröning,~O.; Overbeck,~J.; Braun,~O.; Perrin,~M.~L.; Borin~Barin,~G.;
  El~Abbassi,~M.; Eimre,~K.; Ditler,~E.; Daniels,~C.; Meunier,~V.;
  Pignedoli,~C.~A.; Calame,~M.; Fasel,~R.; Ruffieux,~P. Massive Dirac Fermion
  Behavior in a Low Bandgap Graphene Nanoribbon Near a Topological Phase
  Boundary. \emph{Advanced Materials} \textbf{2020}, \emph{32}, 1906054\relax
\mciteBstWouldAddEndPuncttrue
\mciteSetBstMidEndSepPunct{\mcitedefaultmidpunct}
{\mcitedefaultendpunct}{\mcitedefaultseppunct}\relax
\EndOfBibitem
\bibitem[Shen \latin{et~al.}(2017)Shen, Gao, and Fuchs]{shen2017frontiers}
Shen,~Q.; Gao,~H.-Y.; Fuchs,~H. Frontiers of on-surface synthesis: from
  principles to applications. \emph{Nano Today} \textbf{2017}, \emph{13},
  77--96\relax
\mciteBstWouldAddEndPuncttrue
\mciteSetBstMidEndSepPunct{\mcitedefaultmidpunct}
{\mcitedefaultendpunct}{\mcitedefaultseppunct}\relax
\EndOfBibitem
\bibitem[Liu \latin{et~al.}(2017)Liu, Chen, and Wu]{liu2017surface}
Liu,~J.; Chen,~Q.-W.; Wu,~K. On-surface construction of low-dimensional
  nanostructures with terminal alkynes: Linking strategies and controlling
  methodologies. \emph{Chin. Chem. Let.} \textbf{2017}, \emph{28},
  1631--1639\relax
\mciteBstWouldAddEndPuncttrue
\mciteSetBstMidEndSepPunct{\mcitedefaultmidpunct}
{\mcitedefaultendpunct}{\mcitedefaultseppunct}\relax
\EndOfBibitem
\bibitem[Klappenberger \latin{et~al.}(2015)Klappenberger, Zhang, Bj{\"o}rk,
  Klyatskaya, Ruben, and Barth]{klappenberger2015surface}
Klappenberger,~F.; Zhang,~Y.-Q.; Bj{\"o}rk,~J.; Klyatskaya,~S.; Ruben,~M.;
  Barth,~J.~V. On-surface synthesis of carbon-based scaffolds and nanomaterials
  using terminal alkynes. \emph{Acc. Chem. Res.} \textbf{2015}, \emph{48},
  2140--2150\relax
\mciteBstWouldAddEndPuncttrue
\mciteSetBstMidEndSepPunct{\mcitedefaultmidpunct}
{\mcitedefaultendpunct}{\mcitedefaultseppunct}\relax
\EndOfBibitem
\bibitem[Sun \latin{et~al.}(2018)Sun, Zhang, Qiu, Liu, and Xu]{Sun2018}
Sun,~Q.; Zhang,~R.; Qiu,~J.; Liu,~R.; Xu,~W. On-Surface Synthesis of Carbon
  Nanostructures. \emph{Adv. Mat.} \textbf{2018}, \emph{30}, 1705630\relax
\mciteBstWouldAddEndPuncttrue
\mciteSetBstMidEndSepPunct{\mcitedefaultmidpunct}
{\mcitedefaultendpunct}{\mcitedefaultseppunct}\relax
\EndOfBibitem
\bibitem[Sun \latin{et~al.}({2017})Sun, Tran, Cai, Ma, Yu, Yuan, Stohr, and
  Xu]{Sun_AngewChem_2017}
Sun,~Q.; Tran,~B.~V.; Cai,~L.; Ma,~H.; Yu,~X.; Yuan,~C.; Stohr,~M.; Xu,~W.
  {On-Surface Formation of Cumulene by Dehalogenative Homocoupling of Alkenyl
  gem-Dibromides}. \emph{{Angew. Chem. Int. Ed.}} \textbf{{2017}}, \emph{{56}},
  {12165--12169}\relax
\mciteBstWouldAddEndPuncttrue
\mciteSetBstMidEndSepPunct{\mcitedefaultmidpunct}
{\mcitedefaultendpunct}{\mcitedefaultseppunct}\relax
\EndOfBibitem
\bibitem[Shu \latin{et~al.}({2018})Shu, Liu, Zha, Pan, Zhang, Xie, Chen, Yuan,
  Qiu, and Liu]{Shu_NatComm_2018}
Shu,~C.-H.; Liu,~M.-X.; Zha,~Z.-Q.; Pan,~J.-L.; Zhang,~S.-Z.; Xie,~Y.-L.;
  Chen,~J.-L.; Yuan,~D.-W.; Qiu,~X.-H.; Liu,~P.-N. {On-surface synthesis of
  poly(p-phenylene ethynylene) molecular wires via in situ formation of
  Carbon-Carbon triple bond}. \emph{{Nat. Commun.}} \textbf{{2018}},
  \emph{{9}}, 2322\relax
\mciteBstWouldAddEndPuncttrue
\mciteSetBstMidEndSepPunct{\mcitedefaultmidpunct}
{\mcitedefaultendpunct}{\mcitedefaultseppunct}\relax
\EndOfBibitem
\bibitem[Yang \latin{et~al.}(2018)Yang, Gebhardt, Schaub, Sander,
  Schönamsgruber, Soni, Görling, Kivala, and Maier]{yang2018nanoscale}
Yang,~Z.; Gebhardt,~J.; Schaub,~T.~A.; Sander,~T.; Schönamsgruber,~J.;
  Soni,~H.; Görling,~A.; Kivala,~M.; Maier,~S. Two-dimensional delocalized
  states in organometallic bis-acetylide networks on Ag(111). \emph{Nanoscale}
  \textbf{2018}, \emph{10}, 3769--3776\relax
\mciteBstWouldAddEndPuncttrue
\mciteSetBstMidEndSepPunct{\mcitedefaultmidpunct}
{\mcitedefaultendpunct}{\mcitedefaultseppunct}\relax
\EndOfBibitem
\bibitem[Rabia \latin{et~al.}(2019)Rabia, Tumino, Milani, Russo, Li~Bassi,
  Achilli, Fratesi, Onida, Manini, Sun, Xu, and Casari]{rabia2019scanning}
Rabia,~A.; Tumino,~F.; Milani,~A.; Russo,~V.; Li~Bassi,~A.; Achilli,~S.;
  Fratesi,~G.; Onida,~G.; Manini,~N.; Sun,~Q.; Xu,~W.; Casari,~C.~S. Scanning
  tunneling microscopy and Raman spectroscopy of polymeric sp–sp2 carbon
  atomic wires synthesized on the Au(111) surface. \emph{Nanoscale}
  \textbf{2019}, \emph{11}, 18191--18200\relax
\mciteBstWouldAddEndPuncttrue
\mciteSetBstMidEndSepPunct{\mcitedefaultmidpunct}
{\mcitedefaultendpunct}{\mcitedefaultseppunct}\relax
\EndOfBibitem
\bibitem[Sun \latin{et~al.}(2016)Sun, Cai, Ma, Yuan, and Xu]{sun2016}
Sun,~Q.; Cai,~L.; Ma,~H.; Yuan,~C.; Xu,~W. Dehalogenative homocoupling of
  terminal alkynyl bromides on Au (111): incorporation of acetylenic
  scaffolding into surface nanostructures. \emph{ACS Nano} \textbf{2016},
  \emph{10}, 7023--7030\relax
\mciteBstWouldAddEndPuncttrue
\mciteSetBstMidEndSepPunct{\mcitedefaultmidpunct}
{\mcitedefaultendpunct}{\mcitedefaultseppunct}\relax
\EndOfBibitem
\bibitem[Fratesi \latin{et~al.}(2018)Fratesi, Achilli, Manini, Onida, Baby,
  Ravikumar, Ugolotti, Brivio, Milani, and Casari]{frat2018}
Fratesi,~G.; Achilli,~S.; Manini,~N.; Onida,~G.; Baby,~A.; Ravikumar,~A.;
  Ugolotti,~A.; Brivio,~G.~P.; Milani,~A.; Casari,~C.~S. Fingerprints of sp1
  Hybridized C in the Near-Edge X-ray Absorption Spectra of Surface-Grown
  Materials. \emph{Materials} \textbf{2018}, \emph{11}, 2556\relax
\mciteBstWouldAddEndPuncttrue
\mciteSetBstMidEndSepPunct{\mcitedefaultmidpunct}
{\mcitedefaultendpunct}{\mcitedefaultseppunct}\relax
\EndOfBibitem
\bibitem[Donati \latin{et~al.}(2009)Donati, Sessi, Achilli, Li~Bassi, Passoni,
  Casari, Bottani, Brambilla, Picone, Finazzi, Du`{o}, Trioni, and
  Ciccacci]{Donati2009}
Donati,~F.; Sessi,~P.; Achilli,~S.; Li~Bassi,~A.; Passoni,~M.; Casari,~C.~S.;
  Bottani,~C.~E.; Brambilla,~A.; Picone,~A.; Finazzi,~M.; Du`{o},~L.;
  Trioni,~M.~I.; Ciccacci,~F. Scanning tunneling spectroscopy of the
  Fe(001)$-$p(1x1)0 surface. \emph{Phys. Rev B} \textbf{2009}, \emph{79},
  195430\relax
\mciteBstWouldAddEndPuncttrue
\mciteSetBstMidEndSepPunct{\mcitedefaultmidpunct}
{\mcitedefaultendpunct}{\mcitedefaultseppunct}\relax
\EndOfBibitem
\bibitem[Zhang \latin{et~al.}({2016})Zhang, Wang, Li, Zhao, Tong, Liu, Zhang,
  and Liu]{Zhang_JPCC2016-graphyne}
Zhang,~S.; Wang,~J.; Li,~Z.; Zhao,~R.; Tong,~L.; Liu,~Z.; Zhang,~J.; Liu,~Z.
  {Raman Spectra and Corresponding Strain Effects in Graphyne and Graphdiyne}.
  \emph{{J. Phys. Chem. C}} \textbf{{2016}}, \emph{{120}}, {10111--10720}\relax
\mciteBstWouldAddEndPuncttrue
\mciteSetBstMidEndSepPunct{\mcitedefaultmidpunct}
{\mcitedefaultendpunct}{\mcitedefaultseppunct}\relax
\EndOfBibitem
\bibitem[Fan \latin{et~al.}(2015)Fan, Gottfried, and Zhu]{Fan20152484}
Fan,~Q.; Gottfried,~J.; Zhu,~J. Surface-Catalyzed C-C Covalent Coupling
  Strategies toward the Synthesis of Low-Dimensional Carbon-Based
  Nanostructures. \emph{Acc. Chem. Res.} \textbf{2015}, \emph{48},
  2484--2494\relax
\mciteBstWouldAddEndPuncttrue
\mciteSetBstMidEndSepPunct{\mcitedefaultmidpunct}
{\mcitedefaultendpunct}{\mcitedefaultseppunct}\relax
\EndOfBibitem
\bibitem[Klappenberger \latin{et~al.}(2018)Klappenberger, Hellwig, Du,
  Paintner, Uphoff, Zhang, Lin, Moghanaki, Paszkiewicz, Vobornik,
  \latin{et~al.} others]{klappenberger2018functionalized}
Klappenberger,~F.; Hellwig,~R.; Du,~P.; Paintner,~T.; Uphoff,~M.; Zhang,~L.;
  Lin,~T.; Moghanaki,~B.~A.; Paszkiewicz,~M.; Vobornik,~I., \latin{et~al.}
  Functionalized Graphdiyne Nanowires: On-Surface Synthesis and Assessment of
  Band Structure, Flexibility, and Information Storage Potential. \emph{Small}
  \textbf{2018}, \emph{14}, 1704321\relax
\mciteBstWouldAddEndPuncttrue
\mciteSetBstMidEndSepPunct{\mcitedefaultmidpunct}
{\mcitedefaultendpunct}{\mcitedefaultseppunct}\relax
\EndOfBibitem
\bibitem[Gao \latin{et~al.}(2019)Gao, Liu, Wang, and Zhang]{Gao_2019}
Gao,~X.; Liu,~H.; Wang,~D.; Zhang,~J. Graphdiyne: synthesis{,} properties{,}
  and applications. \emph{Chem. Soc. Rev.} \textbf{2019}, \emph{48},
  908--936\relax
\mciteBstWouldAddEndPuncttrue
\mciteSetBstMidEndSepPunct{\mcitedefaultmidpunct}
{\mcitedefaultendpunct}{\mcitedefaultseppunct}\relax
\EndOfBibitem
\bibitem[Zuo \latin{et~al.}(2019)Zuo, Wang, Zhang, Lu, and Li]{Zuo_2018}
Zuo,~Z.; Wang,~D.; Zhang,~J.; Lu,~F.; Li,~Y. Synthesis and Applications of
  Graphdiyne-Based Metal-Free Catalysts. \emph{Advanced Materials}
  \textbf{2019}, \emph{31}, 1803762\relax
\mciteBstWouldAddEndPuncttrue
\mciteSetBstMidEndSepPunct{\mcitedefaultmidpunct}
{\mcitedefaultendpunct}{\mcitedefaultseppunct}\relax
\EndOfBibitem
\bibitem[Soler \latin{et~al.}(2002)Soler, Artacho, Gale, Garc{\'\i}a, Junquera,
  Ordej\'on, and S\'anchez-Portal]{Sole02}
Soler,~J.~M.; Artacho,~E.; Gale,~J.~D.; Garc{\'\i}a,~A.; Junquera,~J.;
  Ordej\'on,~P.; S\'anchez-Portal,~D. The {SIESTA} method for ab initio
  order-{N} materials simulation. \emph{J. Phys.: Condens. Matter}
  \textbf{2002}, \emph{14}, 2745\relax
\mciteBstWouldAddEndPuncttrue
\mciteSetBstMidEndSepPunct{\mcitedefaultmidpunct}
{\mcitedefaultendpunct}{\mcitedefaultseppunct}\relax
\EndOfBibitem
\bibitem[Perdew \latin{et~al.}(1996)Perdew, Burke, and Ernzerhof]{PBE}
Perdew,~J.~P.; Burke,~K.; Ernzerhof,~M. Generalized Gradient Approximation Made
  Simple. \emph{Phys. Rev. Lett.} \textbf{1996}, \emph{77}, 3865--3868\relax
\mciteBstWouldAddEndPuncttrue
\mciteSetBstMidEndSepPunct{\mcitedefaultmidpunct}
{\mcitedefaultendpunct}{\mcitedefaultseppunct}\relax
\EndOfBibitem
\bibitem[Grimme(2006)]{Grimme}
Grimme,~S. Semiempirical {GGA}-type density functional constructed with a
  long-range dispersion correction. \emph{J. Comput. Chem.} \textbf{2006},
  \emph{27}, 1787--1799\relax
\mciteBstWouldAddEndPuncttrue
\mciteSetBstMidEndSepPunct{\mcitedefaultmidpunct}
{\mcitedefaultendpunct}{\mcitedefaultseppunct}\relax
\EndOfBibitem
\bibitem[Frisch \latin{et~al.}(2016)Frisch, Trucks, Schlegel, Scuseria, Robb,
  Cheeseman, Scalmani, Barone, Petersson, and Nakatsuji]{g09}
Frisch,~M.; Trucks,~G.; Schlegel,~H.; Scuseria,~G.; Robb,~M.; Cheeseman,~J.;
  Scalmani,~G.; Barone,~V.; Petersson,~G.; Nakatsuji,~H. Gaussian Development
  Version, revision I. 13; Gaussian, Inc. \emph{Wallingford, CT} \textbf{2016},
  \relax
\mciteBstWouldAddEndPunctfalse
\mciteSetBstMidEndSepPunct{\mcitedefaultmidpunct}
{}{\mcitedefaultseppunct}\relax
\EndOfBibitem
\bibitem[Figgen \latin{et~al.}(2005)Figgen, Rauhut, Dolg, and Stoll]{ECP_Au}
Figgen,~D.; Rauhut,~G.; Dolg,~M.; Stoll,~H. \emph{Chem. Phys.} \textbf{2005},
  \emph{311}, 227\relax
\mciteBstWouldAddEndPuncttrue
\mciteSetBstMidEndSepPunct{\mcitedefaultmidpunct}
{\mcitedefaultendpunct}{\mcitedefaultseppunct}\relax
\EndOfBibitem
\end{mcitethebibliography}

    \end{document}